\documentclass[aps,amsmath,prd,nofootinbib]{revtex4}
\usepackage{graphicx}
\usepackage{amsmath}
\usepackage{amssymb}
\usepackage{subfigure}
\usepackage{psfrag}
\usepackage{color}

\begin{document}

\title{Computation of the radiation amplitude of oscillons}

\medskip

\author{Gyula Fodor$^1$, P\'eter Forg\'acs$^{1,3}$, Zal\'an Horv\'ath$^2$,
 M\'ark Mezei$^2$}
\affiliation{$^1$MTA RMKI, H-1525 Budapest 114, P.O.Box 49, Hungary,
$^2$Institute for Theoretical Physics, E\"otv\"os University,\\
 H-1117 Budapest, P\'azm\'any P\'eter s\'et\'any 1/A, Hungary,\\
{$^3$LMPT, CNRS-UMR 6083, Universit\'e de Tours, Parc de Grandmont,
37200 Tours, France}
}

\begin{abstract}

  The radiation loss of small amplitude oscillons (very long-living,
  spatially localized, time dependent solutions) in one dimensional
  scalar field theories is computed in the small-amplitude expansion analytically using matched asymptotic series expansions and Borel summation. The amplitude of
  the radiation is beyond all orders in perturbation theory and the
  method used has been developed by Segur and Kruskal in
  Phys.\ Rev.\ Lett.\ {\bf 58}, 747 (1987). Our results are in good
  agreement with those of long time numerical simulations of
  oscillons.

\end{abstract}

\maketitle

\section{Introduction}\label{s:intro}

Time dependent spatially localized solutions in various field
theories, which are long living in the sense of staying localized much
longer than the light crossing time, have been already found in the
seventies \cite{dashen,Kudryav,BogMak2,Makhankov} and ever since they
are still attracting considerable interest
\cite{Geicke3,Gleiser,CopelGM95,Honda}. An obvious reason for interest
is the unexpected longevity of these objects, all of which exhibit
nearly periodic oscillations in time. The presence of at least one
real massive scalar field seems to be necessary in order that such
long living, spatially localized field oscillations -- {\sl oscillons}
-- could form. Since oscillons appear in the course of time evolution
starting from rather generic initial data, this provides another
reason to consider them of physical importance.  Oscillons have been
found to form in physical processes, e.g.\ as a result of
vortex-antivortex annihilation in two dimensional Abelian Higgs model
\cite{GleisThor}, domain wall collapse in $\phi^4$ theory
\cite{Hindmarsh-Salmi07}, QCD phase transition \cite{Kolb:1993hw}, or
during symmetry breaking in three dimensional Abelian Higgs model
\cite{GleiserTHor08}. Therefore there is some reason to believe that
oscillons (or configurations close to them) influence the dynamics,
and they play a r\^ole in phase transitions, cosmology and the
dynamics of extended objects (cosmic strings, domain walls, etc.), see
e.g. \cite{Khlopov,Broadhead:2005hn,Gleiser:2007ts,Borsanyi}.
Importantly oscillons have been also found in the bosonic sector of
the Standard Model \cite{Farhi05,Graham07a,Graham07b}. There are
attempts for including fermion fields in the study of oscillons
\cite{Borsanyi2}.  Oscillons resemble breathers of the one dimensional
sine-Gordon (SG) theory, with the important difference that unlike
true breathers they are continuously loosing energy by slow radiation.
An oscillon just like a breather possesses a localized ``core'', but
differs significantly in its ``radiative'' region outside of the
core. Oscillons have been observed in various spatial dimensions, from
$d=1$ up to $d=6$ \cite{Gleiser04,SafTra}, there is, however, a marked
difference between oscillons in $d\leq2$ and in $d>2$. In dimensions
$d=1$,$2$ oscillons can be well described by an adiabatic time
evolution of breather-like configurations with an ever decreasing
amplitude and with an increasing frequency tending towards a limit
determined by the mass threshold
\cite{PietteZakr98,GleSor,Hindmarsh-Salmi06}. In higher dimensions
there are various types of oscillons, they exhibit instabilities and
their behaviour is more complex \cite{CopelGM95,Honda,FFGR}. Oscillons
have been studied on a $1+1$ dimensional expanding background in Refs.
\cite{GrahStam,fggirs}.

In this paper we consider oscillons in scalar field theories with a
general self-interaction potential in one spatial dimension.  So far
most work on oscillons has been either purely numerical or based on
various approximations. E.g. oscillon energy and lifetime has been
recently estimated in Ref.\ \cite{sicilia}.  Our starting point is the small
amplitude expansion \cite{Kichenassamy,FFHL2} which yields
breather-like configurations with spatially localized cores.
The small amplitude expansion yields an asymptotic series for the core.
Solutions of the field equations are either periodic in time or are radiating.
In the first case standing wave tails are present outside of the breather
core, whereas in the radiative case there are outgoing waves.
These latter correspond to oscillons.
Following Segur and Kruskal \cite{SK} and adapting the approach
developed in Refs.\ \cite{Hakim}, \cite{Pomeau} we compute
analytically one of the most important physical characteristics of
oscillons, the amplitude of the outgoing wave responsible for their
eventual demise. Although strictly speaking our results are only
valid for small amplitude oscillons, this is not a major drawback,
since in $d=1$ the long time behaviour of any oscillon (which is what
we are interested in) is determined by that of the small amplitude
one. The energy loss of a small amplitude oscillon in $d=1$ to leading
order can be written as
\begin{equation}\label{e:energy-loss}
\frac{dE}{dt}=-Ae^{-B/E}\,,
\end{equation}
 where $E$ is the energy of the oscillon core, $A$ and $B$ are constants determined by the
theory. The corresponding equation for the $\phi^4$ theory has been
found first in Ref.\ \cite{SK}. Geicke \cite{geicke1} has verified
numerically the asymptotic energy loss following from Eq.\
\eqref{e:energy-loss}, i.e.\ $E(t)\sim B/\ln t$. Since this is not a
completely straightforward numerical exercise (one has to prepare very
good initial data, one needs long time simulations to high precision,
etc.) we have also made a detailed numerical investigation.  For small
enough amplitudes, $\varepsilon$, one has $E\propto \varepsilon$ and
it is clear that the energy loss is beyond all orders in
$\varepsilon$. In this work we compute the constants $A$, $B$
analytically, using matched asymptotic expansion and Borel summation
techniques, as well as by an independent numerical method.
Also we have been able to give convincing numerical evidence for the validity of the
radiation law \eqref{e:energy-loss} by preparing good initial data and performing long
time simulations. We have also checked the
validity of our analytical and numerical results on the example of the
sine-Gordon theory where $A$ is exactly zero.

The plan of this paper is the following: First, in subsection
\ref{sec:small} we present the essential points of the small amplitude
expansion in a one dimensional scalar field theory and relate it to
the Fourier expansion of time-periodic solutions. Then we analytically
extend the solution of the mode equations to the complex plane.
In subsection \ref{sec:beyond} we relate the radiative tail of oscillons to an exponentially small
correction to the asymptotic expansion.
In subsection \ref{sec:radlaw} the radiation law, Eq.\ \eqref{e:energy-loss}
is derived.
In subsection \ref{ss:comp} we determine the
amplitude of the exponential correction by solving numerically the Fourier mode
equations in the complex plane. In subsection \ref{ss:bs} the Borel
summation procedure is used to calculate this correction analytically.
In second half of the paper numerical simulations are presented
supporting the earlier analytical results.
In subsections \ref{sec:aspects} and \ref{ss:indat} we introduce the methods used in numerical
simulations and for generating good initial data starting from the small amplitude expansion
and employing a tuning respectively. In subsection \ref{sec:reliability} we estimate the
lattice effects by putting the sine-Gordon breather on the lattice and measuring its energy loss.
In subsection \ref{sec:phi4} we investigate $\phi^4$ oscillons evolving from initial data
extracted from the small amplitude expansion.
We found that the functional form
of the theoretical radiation law, Eq. \eqref{e:energy-loss} fits our data points; however,
the fitted value of $A$ differs from the theoretical one.
One reason for the disagreement is the exceptionally small
value of $A$ in the $\phi^4$ theory.
Furthermore we have performed simulations
starting with the initial data of Ref.\ Ref.\ \cite{geicke1}.
In subsection \ref{sec:verification} we investigate oscillon radiation in a
specific $\phi^6$ theory where the value $A$ is maximal.
In this case our numerical results agree with the theoretical one
with satisfactory precision.

\section{Analytic approach to oscillon radiation}

\subsection{Small amplitude expansion \label{sec:small}}

We consider a real scalar theory in a $1+1$ dimensional Minkowski space-time,
with a general self-interaction potential, $U(\phi)$.
The equation of motion is just a non-linear wave equation (NLWE) given by
\begin{equation}\label{e:evol}
 -\phi_{,tt} + \phi_{,xx} = U'(\phi)=\phi +\sum\limits_{k=2}^{\infty}g_k\phi^k\,,
\end{equation}
where $\phi$ is a real scalar field.
In Eq.\ \eqref{e:evol} the mass of the field is chosen to be $1$, and it has
been assumed that the potential, $U(\phi)$, can be written as a power series in
$\phi$ where the $g_k$ are real constants.

As explained in detail in Refs.\ \cite{Kichenassamy,FFHL2} small
amplitude solutions of Eq. \eqref{e:evol} can be represented by the
series:
\begin{equation}
\phi=\sum_{k=1}^\infty\varepsilon^k \phi_k(\tau,\zeta) \,, \label{e:sumphi}
\end{equation}
where $\varepsilon$ is a small parameter and the coordinates have been
rescaled as $\tau=\omega t$ and $\zeta=\varepsilon x$, with
$\omega=\sqrt{1-\varepsilon^2}$. The time dependence of the
functions $\phi_k(\tau,\zeta)$ is found to be determined recursively
by a set of (forced) oscillator equations.  For example the first few $ \phi_k$
can be written as:
\begin{eqnarray}
\phi_1&=&p_1(\zeta)\cos \tau\nonumber\\
\phi_2&=&\frac16 g_2p_1^2(\zeta)\left(\cos(2\tau)-3\right)\label{e:phi3b}\\
\phi_3&=&p_3(\zeta)\cos \tau+\frac{1}{72}(4g_2^2-3\lambda)
p_1^3(\zeta)\cos(3\tau)
\nonumber\,,
\end{eqnarray}
where $p_1(\zeta)$, $p_3(\zeta)$ are given in terms of
the parameters of the potential and of a single
function, $S(\zeta)$ as:
\begin{equation}
p_1(\zeta)=\frac{S(\zeta)}{\sqrt{\lambda}}\,, \qquad
p_3(\zeta)=\frac{1}{\lambda^{5/2}}\left[
\left(\frac{1}{24}\lambda^2-\frac{1}{6}\lambda g_2^2+\frac{5}{8}g_5
-\frac{7}{4}g_2g_4+\frac{35}{27}g_2^4\right)Z(\zeta) -\frac{1}{54}\lambda
g_2^2S(\zeta)(32+19S^2(\zeta)) \right]\,, \label{e:p3}
\end{equation}
together with $Z=S(4-S^2)/3$, and $\lambda=5g_2^2/6-3g_3/4>0$.
The function $S(\zeta)$ is a globally regular solution of the following nonlinear equation:
\begin{equation}\label{e:Sequation}
\frac{{\rm d}^2S}{{\rm d}\zeta^2} -S +S^3=0\,.
\end{equation}
By fixing the center of symmetry at $\zeta=0$ the regular solution of Eq. \eqref{e:Sequation} is
given simply by
\begin{equation}
    S(\zeta)=\sqrt{2}\,{\rm sech}(\zeta)\;.
\end{equation}
The only condition on the potential ensuring
the existence of an exponentially decreasing solution to Eq.\eqref{e:Sequation}
is the positivity of the parameter $\lambda$.
It is clear that each term in the series Eq. \eqref{e:sumphi} is exponentially decaying in space
and is periodic in time, hence if it would converge it would yield exponentially localized
breathers.
It has been demonstrated by Segur and Kruskal \cite{SK} for the
example of the $\phi^4$ theory with $U(\phi)=\phi^2(\phi-2)^2$/8, that
the small amplitude expansion \eqref{e:sumphi} does not converge,
actually it is an asymptotic series. Nevertheless for sufficiently
small values of $\varepsilon$ this asymptotic series constitutes an
excellent approximation for oscillons of frequency
$\omega(\varepsilon)$ over a large interval in $\zeta$
\cite{FFHL2}.  The physical reason for the absence of
spatially localized, exactly time-periodic breathers is simply that
for generic potentials, time-dependent solutions of the NLWE
\eqref{e:evol} radiate. This makes it very plausible that
the asymptoticity of the series \eqref{e:sumphi} in general models is
due to radiative phenomena. An interesting prototype exception,
i.e. when \eqref{e:sumphi} converges, is the celebrated breather
solution of the sine-Gordon (SG) theory, where $U(\phi)=1-\cos(\phi)$.
The SG breather can be written as
\begin{equation}\label{e:sG-breather}
\phi(x,t)_{B}=4\arctan\left[
\frac{\varepsilon\sin(\omega(\varepsilon)t)}{\omega(\varepsilon)\cosh(\varepsilon x)}\right]=
4\arctan\left[
\frac{\varepsilon}{\sqrt{1-\varepsilon^2}}\frac{S(\zeta)\sin\tau}{\sqrt2}\right]\,.
\end{equation}
It is a simple matter now to show that the small amplitude expansion
of the SG breather yields a convergent series, indeed.
Let us note here that in Eq.\ \eqref{e:sG-breather} $|\varepsilon|<1$, and it must not be small.
By performing
the $\varepsilon$ expansion of \eqref{e:sG-breather} we obtain
\begin{equation}\label{e:sG-series}
\phi(\tau,\zeta)_{B}=\sqrt{2}\;\,\varepsilon S(\zeta)\left[2+\varepsilon^2-\varepsilon^2S^2(\zeta)/4\right]
\sin\tau+\sqrt{2}\,\varepsilon^3 S^3(\zeta)\sin(3\tau)/6
+\mathcal{O}(\varepsilon^5)
\end{equation}
and it is easy to see that this series converges for $\vert\varepsilon
S\vert<1$, thus for $\varepsilon<1/{\sqrt{2}}$ the small amplitude
expansion of the SG breather converges.

For a given theory the series \eqref{e:sumphi} is unique.
Since all of its terms are time-periodic, it
represents a family of breather-like solutions in the sense of an asymptotic series.
At this point it is natural to look for time-periodic solutions of the NLWE Eq.\
\eqref{e:evol} by expanding the field $\phi$ in Fourier series:
\begin{equation}
\phi(\zeta,\tau)=\sum_{k=0}^{\infty} \Phi_k(\zeta) \,\cos(k\tau)\;,
\end{equation}
leading to an infinite set of mode equations for the $\Phi_k$
\begin{align}
\left[\varepsilon^2\frac{{\rm d}^2}{{\rm d}\zeta^2} -1\right] \Phi_0
&=g_2\Phi_0^2 + g_3\Phi_0^3 +
 \left(\frac{3g_3}{2}\,\Phi_0+\frac{g_2}{2}\right) \sum_{m=1}^\infty \Phi_m^2
+\frac{g_3}{4}\sum_{m,p,q =1}^{\infty} \Phi_m \Phi_p \Phi_q\,
\delta_{m , \pm p \pm q} + \ldots \nonumber\\
 \left[\varepsilon^2\frac{{\rm d}^2}{{\rm d}\zeta^2} +(n^2
\omega^2-1)\right] \Phi_n &= (3g_3\,\Phi_0^2+2g_2\,\Phi_0)\Phi_n
+\left(\frac{3g_3}{2}\,\Phi_0+\frac{g_2}{2}\right)  \sum_{m,p = 1}^\infty
\Phi_m \Phi_p\,
\delta_{n , \pm m \pm p} \label{e:modes}\\
\nonumber  &+ \frac{g_3}{4}\sum_{m,p,q = 1}^{\infty} \Phi_m \Phi_p \Phi_q\,
\delta_{n , \pm m \pm p \pm q}+\ldots\;, \quad n=1,2,\ldots\;,
\end{align}
where $\delta_{m , \pm p \pm q}=\delta_{m,p+q}+\delta_{m,p-q}
+\delta_{m,-p+q}+\delta_{m,-p-q}$.
A remarkable simplification takes place, if the potential is symmetric
around zero, i.e.\ $g_{2k}=0$ for $k=1,2,\ldots\;$. In this case only the
odd Fourier coefficients are nonzero in the Fourier expansion and the
mode equations take the form:
\begin{equation}
\left[\varepsilon^2\frac{{\rm d}^2}{{\rm d}\zeta^2} +(n^2 \omega^2-1)\right]
\Phi_n
=\frac{g_3}{4}\,\sum_{m,p,q=1}^{\infty}\Phi_m\Phi_p\Phi_q\,\delta_{n,\,\pm m
\pm p\pm q}+\ldots \ , \qquad n, m, p, q=1,3,5\ldots \,.\label{e:modes2}
\end{equation}
Equations \eqref{e:modes} admit solutions with a spatially well
localized core and an oscillating (standing wave) tail whose amplitude
tends to a constant for $\vert x\vert\to\infty$. The asymptotic tail
can be approximated as consisting of a superposition of standing waves
of frequencies $n\omega$, $n=2,3,\ldots\,$. We note that for bounded
solutions all modes $\Phi_n$ for $n>1$ contain two parameters which
can be interpreted as an amplitude and a phase of the corresponding
frequency standing wave.  We are interested in solutions for which the
amplitudes of the tails are much smaller than that of the core, and
these are the ones related to oscillons.  Because of the existence of
the asymptotic standing wave tail these solutions are not unique.
Intuitively it is clear that for a given frequency, solutions with the
smallest possible amplitude standing wave tail should be close to the
``inner part'' of oscillons.  It is quite plausible that for a fixed
value of $\varepsilon$, the solution with a minimal amplitude tail and
being symmetric with respect to the origin, is actually unique.  This
hypothesis is supported by the results of Ref.\ \cite{FFGR}, where
such breather-type solutions have been named quasi-breathers.  Another
type of quasi-breather is the (most likely unique) solution, for
which all modes, $\Phi_n\to 0$ exponentially for say $x\to-\infty$,
with a small amplitude oscillating tail in the other direction. We
shall refer to such objects as asymmetric quasi-breathers.

For sufficiently small values of $\varepsilon$ the tail amplitudes
become exponentially small in $\varepsilon$. In this case the core can
be treated separately from the tail, and a linear superposition of the
two is a very good approximation to the solution.  One can verify that
the small amplitude expansion of \eqref{e:modes} reproduces the terms
of the asymptotic expansion \eqref{e:phi3b}. We remark that the small
amplitude expansion represents an essentially unique core. As already
mentioned the amplitude of the standing wave tail is ${\cal
  O}(\exp(-1/\varepsilon))$, i.e. it is beyond all orders in
perturbation theory.  Segur and Kruskal (SK) \cite{SK} has worked out
a method to compute this ``transcendentally small'' amplitude on the
example of the $\phi^4$ theory using matched asymptotic series
expansion.  In the following we shall use the SK method to find the
amplitude of the standing wave tail of quasi-breathers when
$\varepsilon\to0$.  The main idea of the SK method is to define an
``inner'' problem in the complex $\zeta$-plane in the neighborhood of
the singularity of $S$ closest to the real axis. Clearly it is located
at $\zeta=\pm i\pi/2$ where the function $S$ has a simple pole. In
fact close to the pole at $i\pi/2$
\begin{equation}
S(y)=-\frac{i\,  \sqrt2}{\varepsilon y}+\frac{i\, \sqrt2\, \varepsilon
y}{6}+\mathcal{O}\left((\varepsilon y)^3\right)\;,
\end{equation}
where the rescaled variable $y$ is defined as
\begin{equation}
\zeta=i \pi/2+\varepsilon y \,.
\end{equation}
The inner problem is defined by the rescaled variables near the
singularity and keeping only the leading terms in $\varepsilon$.  For
example, for symmetric potentials the inner equations are:
\begin{equation}
\left[\frac{{\rm d}^2}{{\rm d}y^2} +(n^2 -1)\right] \Phi_n
=\frac{g_3}{4}\,\sum_{m,p,q=1}^{+\infty}\Phi_m\Phi_p\Phi_q\,\delta_{n,\,\pm m
\pm p\pm q}+\ldots \label{e:compmod2}
\end{equation}
where it has been also used that $\omega^2=1$ to leading order in
$\varepsilon$.  We look for such solutions of the mode equations of
the inner problem which can be matched to the solution of the outer problem. In our case the outer problem is defined by the analytic continuation of
the small amplitude expansion from the real axis. The matching region
is parametrized in the following way:
\begin{equation}
\left\{\left|\varepsilon y\right|\ll 1 \;\; (\varepsilon y\rightarrow
0),\;\;\left|y\right|\gg1\;\; (\left|y\right|\rightarrow \infty), \;\;
-\pi\leq\mathrm{arg}(y)\leq -\frac{\pi}{2} \right\} \,. \label{e:matchreg}
\end{equation}
Making use of the fact that the leading $\varepsilon$ order of
$\Phi_n$ on the real axis is proportional to $\varepsilon^n\,S^n$ it
follows that in the matching region
\begin{align}
\Phi_0&=\sum_{k=2}^{+\infty}
\frac{a^{(0)}_k}{y^k}+\mathcal{O}(\varepsilon^2)
\label{e:Phi0}\\
\Phi_n&=\sum_{k=n}^{+\infty} \frac{a^{(n)}_k}{y^k}+\mathcal{O}(\varepsilon^2)
\label{e:Phin}\,
\end{align}
where the coefficients $a^{(n)}_k$ are uniquely determined by the
small amplitude expansion on the real axis.  The first few terms in
the expansion \eqref{e:Phi0}\eqref{e:Phin}, up to order $1/y^4$, can
be written as
\begin{align}
\Phi_0&=\frac{g_2}{\lambda}\,\frac{1}{y^2}+\ldots\nonumber\\
\Phi_1&=-\frac{i \sqrt2}{\sqrt\lambda}\,\frac{1}{y}
-\frac{i 2\sqrt2}{3\lambda^2\sqrt{\lambda}}
\left(\frac{1}{24}\lambda^2+\frac{8}{9}\lambda g_2^2+\frac{5}{8}g_5
-\frac{7}{4}g_2g_4+\frac{35}{27}g_2^4\right)\,\frac{1}{y^3}
+\ldots \label{e:1match} \\
 \Phi_2&=-\frac{g_2}{3\lambda}\,\frac{1}{y^2}+\ldots\nonumber\\
 \Phi_3&=\frac{i \sqrt2\,(4g_2^2-3\lambda)}{36\lambda^{3/2}}\,
\frac{1}{y^3}+\dots\nonumber \,.
\end{align}

\subsection{Correction beyond all orders\label{sec:beyond}}

In this subsection we will construct an exponential correction to the
asymptotic series, Eq. \eqref{e:1match} which after matching to the outer region determines the amplitude of the standing
wave tail.  A correction beyond all orders to a divergent series might
seem meaningless at first glance, however, following Ref.\ \cite{SK}, we can
give meaning to the method by finding a place where at least the
imaginary part of the original series converges.  This region is the
imaginary axis ${\rm Re}\,y=0$, where the algebraic asymptotic series
\eqref{e:1match} is real, so $\mathrm{Im}\Phi_n=0,$ and the imaginary
part of the series converges trivially. We divide $\Phi_n$ into real
and imaginary parts,
\begin{equation}
    \Phi_n=\Psi_n+i\Omega_n,
\end{equation}
and decompose the inner version of Eq. \eqref{e:modes} (with
$\varepsilon=0$ and using the variable $y$) into real and imaginary
parts.  Then one can linearize the imaginary parts of the mode
equations, and obtain coupled linear differential equations for
$\Omega_n$ along the imaginary axis. These equations contain
$\Omega_m$ terms multiplied by various powers of $\Psi_k$. As a first
approximation, in the matching region these can be neglected, and one
gets decoupled homogeneous linear differential equations with constant
coefficients for $\Omega_n$. For $\Omega_0$ the solutions are
$\Omega_0=\exp\left(\pm y\right)$, which are oscillating,
non-decreasing functions in the direction of the imaginary axis and
have to be omitted, as they cannot be matched to the $\varepsilon$
expansion on the real axis. The solutions for $\Omega_1$ are linear in
$y$ and the matching conditions forbid them as in the previous
case. The solutions $\Omega_n$ for $n>1$ which are tending to zero as
${\rm Im}\,y\to-\infty$ are
\begin{equation}
 \Omega_n=\nu_n\,\exp\left(-i\sqrt{n^2-1}\, y\right) \,.
\end{equation}
Because of linearization one cannot determine the amplitudes
$\nu_n$; methods to calculate them will be presented in the next
subsections.

If $\nu_2\neq 0$ the dominant among the exponential corrections is
$\Omega_2$ which will yield the leading term in the radiation.  It is
possible to get correction terms to $\nu_n$ by taking into account
terms proportional to $\Psi_k$ in the differential equations and
substituting them by the leading order terms of the asymptotic
expansion.  Also considering the coupling between different
$\Omega_n$, we get
\begin{equation}
\Omega_2=\nu_2\,\exp\left(-i\sqrt{3}\,
y\right)\left[1+\frac{2i\sqrt{3}(g_2^2-3\lambda)}{9\lambda y}
+\mathcal{O}\left(\frac{1}{y^2}\right)\right]
+\mathcal{O}\left(\frac{1}{y}\,\exp\left(-i\sqrt{8}\,
y\right)\right)\,. \label{e:om2}
\end{equation}
In the case of symmetric potentials $\Phi_{2k}$s are absent, so the dominant
contribution comes from
\begin{equation}
\Omega_3=\nu_3\,\exp\left(-i\sqrt{8}\,
y\right)\left[1-\frac{i}{y\sqrt{2}}
+\mathcal{O}\left(\frac{1}{y^2}\right)\right]
+\mathcal{O}\left(\frac{1}{y^2}\,\exp\left(-i\sqrt{24}\,
y\right)\right)\,. \label{e:om3}
\end{equation}
Similar exponential correction appears in the neighborhood of the
singularity $-i\,\pi/2$. We should still match the correction to the imaginary part, Eq. \eqref{e:om2} to the solution on the
real axis. Hence we linearize the equation of $\Phi_2$,
\eqref{e:modes} about the quasi-breather core and get the following solution:
\begin{equation}
\delta\Phi_2=C\,\sin\left(\sqrt3 x+\alpha\right)\,, \label{e:realaxis}
\end{equation}
where $C$ and $\alpha$ are arbitrary constants.  We analytically
continue $\delta\Phi_2$ to the complex plane. We match it to the
exponential correction \eqref{e:om2} obtained in the inner region around
the pole $i\pi/(2\varepsilon)$, and to the corresponding expression around
$-i\pi/(2\varepsilon)$. This determines $C$ and $\alpha$
\begin{equation}
C=2\nu_2\exp\left(-\frac{\sqrt3 \pi}{2\varepsilon}\right)\,, \quad \alpha=0 \ .
\end{equation}
Neglecting contributions from higher Fourier modes the derivative of
the field $\phi$ in the origin oscillates as
\begin{equation}
\partial_x \phi\vert_{ x=0}=2\nu_2\exp\left(-\frac{\sqrt3 \pi}
{2\varepsilon}\right)
\sqrt{3}\cos\left(2t\right)\,. \label{e:orig}
\end{equation}
For symmetric potentials the above matching procedure works in a completely analogous way. 
The original problem of determining a periodic solution of the field
equation, Eq. \eqref{e:evol} is well posed if we impose boundary conditions. We use the
boundary conditions introduced by Segur and Kruskal in Ref.\ \cite{SK},
namely we require the field to vanish at $x\to -\infty$ and that the
solution remains bounded. These requirements provide sufficient conditions
to make the solution unique. We obtained a time-periodic solution of the
field equation which is asymmetric with respect to $x=0$. It is clearly not a breather, since it has a standing wave tail in $x\to
+\infty$ and in this sense it is weakly localized. 

We can add a transcendentally small standing wave to $\Phi_2$ as it
solves the corresponding linearized mode equation about the asymmetric
quasi-breather (AQB) denoted by $\phi_{AQB}$ and it will give the leading order
transcendental term in the new solution. Therefore one can write a
symmetric configuration $\phi_S$ in the following form:
\begin{equation}
\phi_{S}=\phi_{AQB}+\phi_{st}\,, \label{e:SQB}
\end{equation}
where $\phi_{st}$ denotes a transcendentally small standing
wave. $\phi_{st}$ has to make $\phi_{S}$ symmetric about the
origin. This requirement is enough to uniquely determine it. The
derivative of $\phi_{AQB}$ at $x=0$ is given by \eqref{e:orig}, hence
\begin{equation}
\phi_{st}=2\nu_2\exp\left(-\frac{\sqrt3\pi}{2\varepsilon}\right)
\frac{\cos\left(\sqrt{3}\;x+\xi\right)}{\sin \xi}\cos\left(2t\right) \,, \label{e:phist}
\end{equation}
where $\xi$ is an arbitrary constant. The symmetric configuration with
the minimal amplitude tail, corresponding to $\xi=\pi/2$, has been
named quasi-breather in Ref.\ \cite{FFHL2}.  This result is consistent with
our knowledge about symmetric solutions, as when we take into account the standing
wave tail in one mode, one of the two free parameters is required for
symmetrization. From Eq. \eqref{e:phist} one immediately gets the
standing wave tail of the asymmetric quasi-breather:
\begin{equation}
\phi_{AQB}=4\nu_2\exp\left(-\frac{\sqrt3\pi}{2\varepsilon}\right)
\sin\left(\sqrt{3}\;x\right)\cos\left(2t\right)\,\quad
\text{for $x\to +\infty$}. \label{e:AQB}
\end{equation}
In conclusion we were able to determine the asymptotic
field of the asymmetric and symmetric quasi-breathers up to one parameter $\nu_2$
($\nu_3$). In subsections \ref{ss:comp} and \ref{ss:bs} we will
determine this parameter.

\subsection{Radiation law for small amplitude oscillons\label{sec:radlaw}}

We can construct a symmetric time dependent solution of the field
equation by repeating the same steps as in the time periodic case and
replacing the standing wave $\phi_{st}$ with a moving wave
$\phi_{rad}$. Let us denote the oscillon field with $\phi_{osc}$. Then
from $\phi_{osc}=\phi_{AQB}+\phi_{rad}$ it follows that
\begin{equation}
\phi_{rad}=-2\nu_2\exp\left(-\frac{\sqrt3\pi}{2\varepsilon}\right)
\sin\left(\sqrt{3}\;x+2t\right)\,. \label{e:phirad}
\end{equation}
We find that the amplitude of the tail of the quasi-breather
determined by $\xi=\pi/2$  is equal to the amplitude of the
outgoing radiation from the corresponding oscillon. This property has
been already noted in Ref.\ \cite{FFGR}.

After the little digression on quasi-breather tails and the determination of the oscillon radiation field we focus on the radiation law. The asymptotic
oscillon field from Eq. \eqref{e:phirad} is:
\begin{align}
\phi_{osc}&=2\nu_n\exp\left(-\frac{\pi\sqrt{n^2-1}}
{2\varepsilon}\right)\sin\left(\sqrt{n^2-1}\;x-nt\right)
& \text{for $x\to +\infty$,}\\
\phi_{osc}&=-2\nu_n\exp\left(-\frac{\pi\sqrt{n^2-1} }{2\varepsilon}\right)\sin\left(\sqrt{n^2-1}\;x+nt\right)
& \text{for $x\to -\infty$,}
\end{align}
where $n=2$ for asymmetric potentials and $n=3$ for symmetric potentials.

The energy carried away by these oscillating tails determines the
time-averaged radiation power $W$ of the oscillon:
\begin{equation}
\frac{{\rm d}E}{{\rm d}t}=W  =-4n\sqrt{n^2-1}\,\nu_n^2\,
\exp\left(-\frac{\pi\sqrt{n^2-1}}{\varepsilon}\right)\;. \label{e:radlow1}
\end{equation}
Since the radiation field is transcendentally small, it is reasonable to assume
that during its time-evolution the core of the oscillon goes through
undistorted quasi-breather states. This statement will be referred to as the
adiabatic hypothesis.
The energy content $E$ of the core of the quasi-breathers as a function of $\varepsilon$
(or equivalently as a function of its frequency $\omega$) is easily determined, it is given
by:
\begin{equation}
E=\frac{2\varepsilon}{\lambda}+{\cal O}(\varepsilon^3)\,. \label{e:energy}
\end{equation}
Now the equation determining the change of the core energy with time (energy loss)
for oscillons can be seen to be of the form given by
Eq.\ \eqref{e:energy-loss}.
From Eq.\ \eqref{e:energy-loss} one can easily deduce that the leading order
late time behaviour of the energy is given as
\begin{equation}\label{e:energy-time}
E(t)\approx\frac{B}{\ln t}\left(1-\frac{2\ln\ln t}{\ln t}\right)
\end{equation}
In this subsection the radiation law for small amplitude oscillons has been
determined up to a single unknown parameter, $A$. The
problem of finding $A$ or what is equivalent the parameters $\nu_2$ resp.\ $\nu_3$ will be done in the
following two subsections.

\subsection{Determining the radiation amplitude by solving the complex
  mode equations numerically}\label{ss:comp}

In this subsection we numerically determine the leading radiation
amplitude coefficients, namely $\nu_2$ for asymmetric potentials and
$\nu_3$ in case of symmetric potentials.  First, we consider the
$\phi^4$ theory, in which case the only non-vanishing coefficients in
the expansion of the potential are $g_2=-\frac{3}{2}$ and
$g_3=\frac{1}{2}$.  We consider various order truncations of the
Fourier mode equations \eqref{e:modes} in the region close to the
singularity. To illustrate our method we present in more details the
calculations for the simplest truncated system that radiates,
i.e. that for which only up to $\cos(2\tau)$ modes are kept. For the
inner problem, in the $\varepsilon\to 0$ limit, the mode equations
\eqref{e:modes} are
\begin{eqnarray}
\frac{\partial^2\Phi_0}{\partial y^2}-\Phi_0&=&\frac{1}{2}\Phi_0^3
-\frac{3}{2}\Phi_0^2
+\frac{3}{4}\left(\Phi_1^2+\Phi_2^2\right)\left(\Phi_0-1\right)
+\frac{3}{8}\Phi_1^2\Phi_2 \ ,\nonumber\\
\frac{\partial^2\Phi_1}{\partial y^2}&=&\frac{3}{2}\Phi_0^2\Phi_1
+\frac{3}{4}\Phi_1\left(\Phi_2-2\right)\left(2\Phi_0+\Phi_2\right)
+\frac{3}{8}\Phi_1^3\ ,\label{phi4modes}\\
\frac{\partial^2\Phi_2}{\partial y^2}+3\Phi_2&=&
\frac{3}{2}\Phi_0^2\Phi_2
+\frac{3}{4}\Phi_1^2\left(\Phi_0+\Phi_2-1\right)
-3\Phi_0\Phi_2+\frac{3}{8}\Phi_2^3 \ . \nonumber
\end{eqnarray}
Expanding the mode equations into powers of $1/y$
\begin{equation}
\Phi_i=\sum_{j=1}^{\infty}a^{(i)}_{j}\,\frac{1}{y^{j}} \ ,
\end{equation}
we find that all coefficients are fixed after choosing $a^{(1)}_{2}$
and the sign in $a^{(1)}_{1}=\pm 2i/\sqrt{3}$.  Furthermore,
$a^{(0)}_{j}=0$, for $j<2$ and $a_i^{(j)}=0$ for $j<i$.  In order
to agree as well as possible with \eqref{e:1match} obtained from the
$\varepsilon$ expansion evaluated at the singularity region we set
$a^{(1)}_{2}=0$ and $a^{(1)}_{1}=-2i/\sqrt{3}$.  Then it turns out
that $a^{(i)}_{i+1+2j}=0$ for all integer j. Raising the truncation
order high enough by adding more Fourier modes to the system all
coefficients will necessarily agree with  \eqref{e:1match}. The
expansion consistent with (\ref{phi4modes}) is
\begin{eqnarray}
\Phi_0&=&-\frac{1}{y^2}-\frac{131}{12y^4}-\frac{461381}{1728y^6}
-\frac{631478123}{51840y^8}+{\cal O}\left(\frac{1}{y^{10}}\right) \ ,
\nonumber\\
\Phi_1&=&-\frac{i}{\sqrt{3}}\left[\frac{2}{y}+\frac{17}{3y^3}
+\frac{103877}{1728y^5}+\frac{8278867}{5760y^7}
+{\cal O}\left(\frac{1}{y^{9}}\right)\right] \ ,\label{phi4exp}\\
\Phi_2&=&\frac{1}{3y^2}+\frac{16}{9y^4}+\frac{107597}{5184y^6}
+\frac{19189237}{38880y^8}+{\cal O}\left(\frac{1}{y^{10}}\right)
 \ . \nonumber
\end{eqnarray}

According to expansion (\ref{phi4exp}) the imaginary part
$\Omega_2={\rm Im}\,\Phi_2$ vanishes to all orders on the imaginary
axis. However, this is not a convergent but an asymptotic series, and
the actual solution of (\ref{phi4modes}) may include an exponentially
small correction to $\Omega_2$ on the imaginary axis, in accordance
with \eqref{e:om2}. Of course, the value of $\Omega_2$ will depend on the
chosen boundary conditions. The method introduced by Kruskal and Segur
\cite{SK} is to integrate the differential equations numerically along a
constant ${\rm Im}\,y=y_{i}$ line from a large negative ${\rm Re}\, y=y_{r}$
value to the axis ${\rm Re}\,y=0$. The boundary conditions at $y=y_r+i y_i$
are given by the expansion (\ref{phi4exp}) and its derivative,
truncated to an appropriate order in $1/y$. This works well for modes
$\Phi_i$ with $i\geq 1$, but $\Phi_0$ has the tendency to
exponentially blow up along constant ${\rm Im}\,y$ lines. This numerically
problematic issue can be avoided by treating the equation for $\Phi_0$
as a two point boundary value problem, setting $\Phi_0=0$ at the axis
point $y=iy_i$ and using the asymptotic series as the boundary value
at $y=y_r+i y_i$. The other modes are treated as initial value
problems by specifying their value and first derivatives at $y=y_r+i
y_i$. We note that in case of symmetric potentials $\Phi_{2i}=0$
everywhere, and the integration procedure simplifies to pure initial
value problem. In that case the first radiating mode is $\Phi_3$.

 For the actual numerical integration of the $\phi^4$ system we have
chosen various $y_i$ values in the interval $[-5,-13]$, $y_r$ in
$[-20,-500]$, and the expansion in the initial data was truncated to
orders from $4$ to $20$. The equations were generated by Maple and its
default boundary value problem differential equation solver was used
for the numerical integration. For $y_r<-50$ the obtained values for
$\Omega_2$ were only changing in the fourth digits on varying
$y_r$. The choice of truncation order in $1/y$ in the initial data was
not changing the results in their less than fourth digits if the order
was chosen larger than $8$. In Table \ref{t:imy}. we present the
obtained $y_i$ dependence of the imaginary part $\Omega_2$, when
$y_r=-300$ and the initial data is of order $15$. According to
\eqref{e:om2} we approximate $\nu_2$ to leading order as
$\nu_2^{(0)}=\Omega_2\exp\left(\sqrt{3}y_i\right)$. The $1/y$
correction gives a more precise result
$\nu_2^{(1)}=\Omega_2\exp\left(\sqrt{3}y_i\right)/(1-1/(y_i\sqrt{3}))$.
\begin{table}[htbp]
\begin{tabular}{|c|c|c|c|c|c|}
\hline
$y_i$ & $\Omega_2$ & $\nu_2^{(0)}$ & $\nu_2^{(1)}$, $\nu_2^{(2)}$
 & $\nu_2^{(4)}$ & $\nu_2^{(6)}$\\
\hline
$-4$  &  $9.58187\cdot10^{-6}$ & $0.00977982$ & $0.00854627$
& $0.00885621$ & $0.00858961$ \\ 
$-6$  &  $2.86915\cdot10^{-7}$ & $0.00935565$ & $0.00853442$
& $0.00863027$ & $0.00859166$ \\ 
$-8$  &  $8.74658\cdot10^{-9}$ & $0.00911169$ & $0.00849838$
& $0.00853980$ & $0.00853022$ \\ 
$-10$ & $2.69068\cdot10^{-10}$ & $0.00895494$ & $0.00846614$
& $0.00848769$ & $0.00848446$ \\ 
$-12$ & $8.31980\cdot10^{-12}$ & $0.00884615$ & $0.00844008$
& $0.00845283$ & $0.00845151$ \\ \hline
\end{tabular}
\caption{\label{t:imy}
  Dependence of the exponential correction $\Omega_2$ on the location of the
  integration line ${\rm Im}\,y=y_i$ for the $\phi^4$ theory truncated at
  $\cos(2\tau)$.
  The corresponding radiation amplitudes $\nu_2^{(n)}$ are also given
  by taking into account $n$-th order corrections in $1/y$.
}
\end{table}
We note that for the special case of the $\phi^4$ theory the
coefficient of the $1/y^2$ correction in $\Omega_2$ vanishes, and
hence $\nu_2^{(1)}$ is also valid to second order. For larger values
of $|y_i|$ it is possible to improve the precision by adding even
higher order corrections.

From Table \ref{t:imy}. we may give a first estimate on the actual
value of the radiation amplitude as
$\nu_2=(8.45\pm0.03)\cdot10^{-3}$. However this value turns out to
change drastically when adding higher Fourier modes to the system
(\ref{phi4modes}). In Table \ref{t:phi4modes}. we give the calculated
values for $\nu_2$ when keeping Fourier modes up to order
$\cos(n\tau)$.
\begin{table}[htbp]
\begin{tabular}{|c|c|}
\hline
$n$ & $\nu_2$\\
\hline
$2$  &  $8.45\cdot10^{-3}$
\\ 
$3$  &  $-7.115\cdot10^{-3}$
\\ 
$4$  &  $-8.431\cdot10^{-3}$
\\ 
$5$  &  $-8.454\cdot10^{-3}$
\\ 
$6$  &  $-8.454\cdot10^{-3}$
\\ \hline
\end{tabular}
\caption{\label{t:phi4modes}
Dependence of the radiation amplitude on the order
of the used Fourier expansions in the $\phi^4$ theory.
}
\end{table}
It is somewhat surprising that the addition of the $\cos(3\tau)$ mode
changes the sign of $\nu_2$, while the magnitude is quite close to the
proper value. The addition of higher than $\Phi_6$ modes does not make
any significant change in the value of $\nu_2$. As a conclusion, we
can state that $\nu_2=(-8.454\pm0.01)\cdot10^{-3}$. The numerical value
obtained by Kruskal and Segur differs by a factor of $2$ due to their use of complex notations for the Fourier modes.
In our units their result is
$(-9.0\pm2.0)\cdot10^{-3}$.

In case of symmetric potentials $\Phi_i=0$ for even $i$, and the
numerical integration method simplifies considerably. A specific
example we have looked at is a specific $\phi^6$ theory, with
$U'(\phi)=\phi-\phi^3+\phi^5$. When using the mode equations only for
$\Phi_1$ and $\Phi_3$ we get $\nu_3=-0.91026$. Adding the fifth and
seventh modes changes the result to $-0.90982$ and $-0.90977$,
respectively. Since the addition of even higher modes do not make
significant change, we can state that for the $\phi^6$ theory:
\begin{equation}
\nu_3=-0.9098\pm0.0001\,. \label{e:phi6nu3}
\end{equation}
An important check of the reliability of our method is to calculate
$\nu_3$ for the sine-Gordon potential $U(\phi)=1-\cos(\phi)$. In this
case, when taking into account all the mode equations and all
expansion coefficients in the expansion of the potential, the exact
result is known to be zero. Instead of changing these independently,
when we solved mode equations up to order $\Phi_i$ we assumed that
$g_j=0$ for $j>i$. As it can be seen from Table \ref{t:sinegordon}.,
the results appear to tend to zero fast as $i$ increases.
\begin{table}[htbp]
\begin{tabular}{|c|c|}
\hline
$i$ & $\nu_3$\\
\hline
$4$  &  $2.32$
\\ 
$6$  &  $-0.2316$
\\ 
$8$  &  $8.35\cdot10^{-3}$
\\ 
$10$  &  $1.14\cdot10^{-5}$
\\ 
$12$  &  $7.2\cdot10^{-7}$
\\ \hline
\end{tabular}
\caption{\label{t:sinegordon}
Radiation amplitude $\nu_3$ for the sine-Gordon theory truncated to
order $i$ in both the mode equations and potential expansion.
}
\end{table}

\subsection{Determining the radiation amplitude by Borel summation of the
algebraic asymptotic series}\label{ss:bs}

In this subsection we will solve Eq. \eqref{e:compmod2} using the
algebraic asymptotic series ansatz \eqref{e:Phin} in the neighborhood
of the singularity. Our considerations will be applicable to the case
of symmetric potentials. At the end of this subsection we will briefly
discuss the problem of asymmetric potentials. The solution is unique
as we have to match it to the original asymmetric quasi-breather
 continued analytically from the real axis. We
truncate both the Taylor expansion of the potential and the Fourier
expansion in order to have a finite set of equations and will work
until we reach convergence as in subsection \ref{ss:comp}.

We will demonstrate how the determination of $a^{(n)}_k$ coefficients
works for big $k$ in leading order of $k$ in the case of minimal
system $\Phi_1$ and $\Phi_3$ with a cubic nonlinearity. Since
$a^{(1)}_k$ and $a^{(3)}_k$ are vanishing for even $k$ we redefine the
coefficients in order to get a more convenient form:
\begin{align}
\Phi_1&=i\,\sum_{k=1}^{\infty} \frac{A_k}{y^{2k-1}}\\
\Phi_3&=i\,\sum_{k=2}^{\infty} \frac{B_k}{y^{2k-1}}\ .
\end{align}
We will show in the following that the behaviour consistent with
Eq. \eqref{e:compmod2} to leading order in $k$ is:
\begin{align}
A_k&\sim \frac{B_k}{2k^2} \label{e:Aas} \\
B_k&\sim K\,(-1)^k\,\frac{(2k-2)!}{8^{k-1/2}} \ . \label{e:Bas}
\end{align}
The constant $K$ can be determined by solving the equations up to some
large order $k$ and matching the gained coefficients to the determined
asymptotic behaviour. To do so, we write up the mode equations:
\begin{align}
\frac{{\rm d}^2}{{\rm d}y^2} \Phi_1& =\frac{3g_3}{4}\Phi_1
\left(\Phi_1^2+\Phi_1\Phi_3+2\Phi_3^2\right) \label{e:bm1}\\
\left(\frac{{\rm d}^2}{{\rm d}y^2} +8\right)
\Phi_3&=\frac{g_3}{4}\left(\Phi_1^3+6\Phi_1^2\Phi_3+3\Phi_3^3\right)\,,
\end{align}
and then determine the equations for the coefficients of
$1/y^{2k-1}$ keeping only terms of order $(2k-4)!$:
\begin{align}
(2k-3)(2k-2)\,A_{k-1}+\frac{3g_3}{4}\,A_1^2 B_{k-1}& =0 \label{e:as1}\\
(2k-3)(2k-2)\,B_{k-1}+8B_k + \frac{3g_3}{2}\,A_1^2 B_{k-1}&=0\;. \label{e:as3}
\end{align}
It is easy to figure out the value of $A_1$ from the matching
conditions \eqref{e:1match}, which gives
$A_1=-\sqrt{-8/(3g_3)}$. (Obviously we get the same result with an
indeterminate sign by solving Eq. \eqref{e:bm1} for the coefficient of
$1/y^3$.) Using the value of $A_1$ we get the asymptotic behaviour of
$A_k$ from Eq. \eqref{e:as1} already given in Eq. \eqref{e:Aas}. From
Eq.  \eqref{e:as3} we can even determine the
$\mathcal{O}\left(1/k^2\right)$ corrections to the asymptotic
behaviour of $B_k$ for large $k$ as
\begin{equation}
B_k\sim
K(-1)^k\frac{(2k-2)!}{8^{k-1/2}}\left(1+\frac{1}{k}+\frac{5}{4k^2}\right)\;.
\end{equation}
This formula enables us to determine the numerical value of $K$ very precisely
from $B_k$s with moderate $k$ value.

By examining the structure of the equations and making use of the fact that the
algebraic asymptotic series of $\Phi_n$ starts only at $1/y^n$ it is
easy to prove that the above asymptotics do not change. Even the
$\mathcal{O}\left(\frac{1}{k^2}\right)$ corrections to $B_k$ are not affected
by the involvement of further modes or higher order nonlinearities. The only
effect of the introduction of further Fourier modes and higher order
nonlinearities which originate in the self-interaction potential is the
changing of the value $K$.

On the one hand, this result gives the proof of the asymptoticity of the series,
i.e. the coefficients are not more gravely divergent than $(2k-2)!$. On the
other hand, this property allows us to Borel-sum the series. It will turn out
that the behaviour of the Borel-summed series in the vicinity of its
singularity gives us the dominant radiation field configuration, i.e. the
exponentially small imaginary correction to $\Phi_3$ on the imaginary axis, in
the matching region.

The first step in the Borel summation is
\begin{align}
  V(z)&=\sum_{k=1}^{\infty}\frac{i\,B_k}{(2k-1)!}\;
  z^{2k-1}\sim\sum_{k=1}^{\infty}iK\,\frac{(-1)^k}{2k-1}\;
  \left(\frac{z}{\sqrt8}\right)^{2k-1}=-\frac{K}{2}\ln\left[\frac{1+iz/\sqrt8}{1-iz/\sqrt 8}\right]\;. \label{e:V}
\end{align}
This Borel summed series has logarithmic singularities at $z=\pm
i\sqrt8$.  The Laplace transform of $V(z)$ will give us the Borel
summed series of $\Phi_3$ which we denote by $\widetilde{\Phi}_3$
\begin{equation}
\widetilde{\Phi}_3(y)=\int_{0}^{\infty} \;\mathrm{d}t\, e^{-t}V\left(\frac{t}{y}\right)\,.\label{borelint}
\end{equation}
The integrand of Eq. \eqref{borelint} has logarithmic singularities at
$t/y=\pm i\sqrt8$. It has been explained in Ref.\ \cite{Pomeau} how to
compute the integral. We only take into account the singularity
$t/y=i\sqrt8$ and not $t/y=-i\sqrt8$, because we would like to get the
correction in the imaginary part of $\Phi_3$ for $y$ points with
negative imaginary parts, as we aim to approach the real axis. For
${\rm Im}\,y<0$ the other singularity stays away from the integration
path, while the singularity from $t/y=i\sqrt8$ appears for
$t=i\sqrt8\,y$. In order to define the integral for ${\rm Re}\,y=0$ we
use the analyticity of $\widetilde{\Phi}_3$. When $y$ is in the
matching region determined by Eq. \eqref{e:matchreg} the singularity
in $t$ is in the lower half-plane, i.e. the contour in
Eq. \eqref{borelint} is above the singularity. When ${\rm Re}\,y\to 0$
the singularity tends to the real axis and the contour must stay above
the singularity. The logaritmic singularity of $V\left(t/y\right)$ does not contribute to the integral and integrating on the branch cut starting from it yields the imaginary
part
\begin{equation}
{\rm Im}\,\tilde{\Phi}_3(y)=\int_{i\sqrt8\,y}^{\infty} \;\mathrm{d}t\; e^{-t}
\,\frac{i\,K\pi}{2}=\frac{i\,K\pi}{2}\,\exp\left[-i\sqrt8\,y\right]\,.
\label{e:borel}
\end{equation}
This result agrees with that of Eq. \eqref{e:om3}, hence we were able to
determine $\nu_3$ analytically:
\begin{equation}
    \nu_3=\frac{K\pi}{2}\;.
\end{equation}

We already gave the method for determining the value of $K$ by solving
linear equations recursively for the coefficients of the algebraic
asymptotic series up to some large $k$ values. In the first step, we
will show that our method is consistent with the fact that the
sine-Gordon breather does not radiate. We will truncate the SG
potential in various orders and determine the value of $K$ by solving
the complex mode equations with the algebraic asymptotic series
ansatz.  We will experience a monotonous decrease in the value of $K$
as we increase the order of the Taylor expansion. This can be
interpreted in the following way.  $K$ determines the increase in the
coefficients of the algebraic asymptotic series which correspond to
the coefficients appearing in the small amplitude expansion. As we get
closer to the SG theory the asymptotic series for small amplitude
quasi-breathers in the truncated SG theory is less and less
growing. On the other hand, this means that the corresponding
oscillons would radiate slower, as the radiation amplitude is
proportional to $K$. In the limit of SG theory we should get a
convergent series in the small amplitude expansion and thus a
non-radiating breather. Our numerical experiences show that we can
even use a smaller number of mode equations than the biggest power in
the Taylor expansion to reach satisfactory convergence for the value
of $K$ for the given theory. We collected the results for $K$ in the
truncated SG theory in Table \ref{t:sGK}.
\begin{center}
\begin{table}[!ht]
\vspace{0.5cm}
\begin{center}
\begin{tabular}{|c|c|c|}
\hline
Order of truncation  & $K$  & $K_{\textbf{num}}$ \\
\hline
4 & $1.486$ & $1.48$\\
6 & $-0.1472$ & $-0.147$\\
8 & $5.306\cdot10^{-3}$ & $5.31\cdot10^{-3}$\\
10 & $7.306\cdot10^{-5}$ & $7.2\cdot10^{-5}$\\
12 & $4.686\cdot10^{-7}$ & $4.6\cdot10^{-7}$\\
\hline
\end{tabular}
\caption{The value of $K$ in truncated SG theories from the solution of the
same number of mode equation as the order of truncation. $K$ denotes
the value obtained by Borel summation, while $K_{\textbf{num}}$ is the
result of the numerical solution of the mode equations on the complex plane.
}\label{t:sGK}
\end{center}
\end{table}
\end{center}

\begin{figure}[!ht]
    \begin{center}
    \includegraphics[width=10cm]{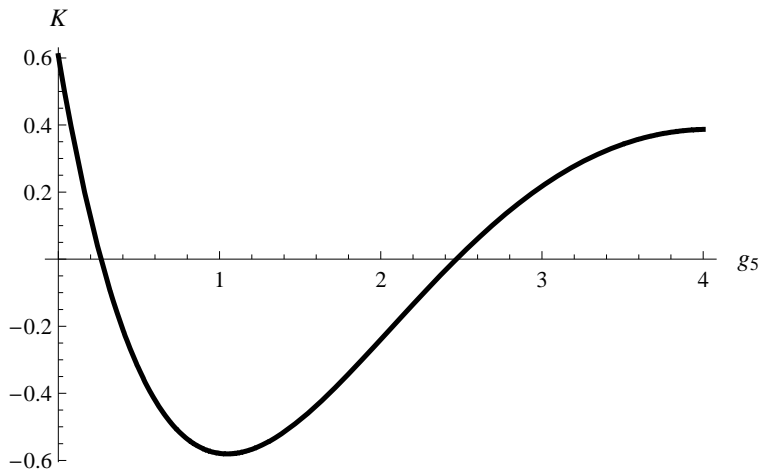}
    \caption{The value of $K$ as a function of $g_5$ for symmetric
      $\phi^6$ theories\label{f:phi6K}}
    \end{center}
\end{figure}

In the second step will will be looking for a symmetric $\phi^6$ theory in
which oscillon radiation is the fastest, i.e. $K$ takes the biggest value. We
will use this theory in numerical simulations for the verification of the
theoretical radiation law for oscillons  because we hope to measure radiation
rate in this theory the most accurately. We will make use of the fact that
$g_3$ can be defined into the fields, thus the only essential parameter in a
symmetric $\phi^6$ theory is $g_5$. ($g_3<0$ in order to have a localized solution of Eq. \eqref{e:Sequation}.) To do so, we solved the theory for
$\Phi_1$-$\Phi_7$ up to $k=30$ with the coefficient $g_5$:
\begin{align}
U(\phi)&=\frac{1}{2}\,\phi^2-\frac{1}{4}\,\phi^4+\frac{g_5}{6}\,\phi^6\\
U'(\phi)&=\phi -\phi^3+g_5\,\phi^5\,.
\end{align}
The value of $K$ as a function of $g_5$ can be found in Fig. \ref{f:phi6K}. We
calculated $K$ form $B_{30}$ in the figure, however the exact value of $k$ only
matters in the sixth digit. We only deal with positive $g_5$s as they are the
ones with a stable vacuum. We see two zeros in this domain. This does not mean
that we found breathers in the corresponding theories, these points represent
oscillons the dominant radiation field of which is in fifth mode, $\Phi_5$.
These configurations are extremely long living objects. The first zero is very
close to the SG theory for which $g_5=3/10$ with the $g_3=-1$ normalization.
Upon introducing further nonlinearities this zero would move exactly to the
$g_k$ point representing the SG theory. From Fig. \ref{f:phi6K} $g_5=1$ seems to be a
very comfortable choice for numerical simulation. We find the precise value of $K$ for $g_5=1$ to be:
\begin{equation}
    K=-0.57915 \ ,\ \  K_{\textbf{num}}=-0.5792 \,,
\end{equation}
where $K_{\textbf{num}}$ has been extracted from Eq. \eqref{e:phi6nu3}.

In the end we will briefly discuss the case of asymmetric potentials the class to which the $\phi^4$ belongs. In these theories the presence of the $\Phi_0$ mode results in the following asymptotics for large $k$s in the $\Phi_0$ mode:
\begin{equation}
	a^{(0)}_k\sim (2k-2)!\;.
\end{equation}
This behaviour dominates the asymptotics of all other modes as well. Because we do not have an alternating sign in this dominant behaviour, the Borel summed series will have a singularity on the real axis, hence we do not get an imaginary correction to the asymptotic series from this calculation. Thus, we cannot determine the radiation this way. The asymptotics which would determine the radiation in the case of asymmetric theories is $a^{(2)}_k\sim (-1)^k\,(2k-2)!/3^k$. This is not the leading behaviour of $\Phi_2$'s coefficients and is significantly suppressed. We have not yet succeeded in determining the hidden alternating part of the coefficients, therefore we cannot determine the radiation amplitude by this analytic method.

\section{Numerical simulations}

\subsection{Aspects of the numerical simulations\label{sec:aspects}}

The numerical simulation of oscillons and their decay were performed
with a slightly modified version of the fourth order method of line
code developed and used in Refs.\ \cite{FR,FR2,FFGR}. The spatial grid was
chosen to be uniform in the compactified radial coordinate $R$ defined
by
\begin{equation}
x=\frac{2R}{\kappa(1-R^2)} \,,
\end{equation}
where $\kappa$ is a constant. By this choice the whole range $0\leq
x<\infty$ of the physical radial coordinate $x$ is mapped to the
interval $0\leq R<1$, avoiding the need for explicitly describing
boundary conditions at some large but finite radius. In most of our
simulations we worked with $\kappa=0.05$, which proved to be ideal for
oscillons for which radiation was numerically observable.

The frequency of oscillons is determined by measuring the time
elapsed between two subsequent maximums of the field configurations in
the origin. By integrating the energy density in an
$R=0.873\;(x=146.7)$ sphere at every timeslice we determine the energy
$E$ of the oscillon core and trace the energy loss $W=dE/dt$.
Although this way of calculating the energy contains an arbitrarity,
for small amplitude oscillons the corresponding asymptotic field is
exponentially small and the energy density of the radiation tail can
be neglected compared to that of the core.  We measure the radiation
in a relatively short time interval from the slope of the energy as a
function of time.

Starting from initial data obtained from the small amplitude expansion
we simulate the time evolution and radiation of oscillons. Although
our analytic calculations are only valid for infinitesimal
$\varepsilon$, we find that they approximate oscillons quite well
after a one-parameter tuning.

Studying the slow energy loss of oscillons we determine a
semi-empirical radiation law. This means that we keep the functional
form \eqref{e:radlow1} but fit the parameters
appearing in it.  We find satisfactory agreement with the theoretical
value of the parameters obtained in the previous sections.  We also
follow the evolution of a single oscillon through a very long time
interval. This process will be precisely described by the radiation obtained by the fit
law proving the assumption made when making analytic considerations:
the system evolves through undistorted quasi-breather states
adiabatically.

\subsection{Initial data}\label{ss:indat}

We used the asymptotic series \eqref{e:sumphi} truncated to order
$N$, at the moment of time-reflection symmetry as initial data for the
numerical evolution code,
\begin{equation} 
\phi^{(\tau=0)}=\sum_{k=1}^{N}\varepsilon^k\phi^{(\tau=0)}_k \,. 
\label{e:indat}
\end{equation}
The aim of our numerical analysis is to obtain oscillon states which
are as periodic as possible, which means that their basic oscillation
frequency and amplitude changes very slowly, uniformly and
monotonically. The time evolution of these ``clean'' oscillons can be
approximated by adiabatic evolution through corresponding frequency
quasi-breather states.  However, since the expansion \eqref{e:sumphi}
is not convergent, the initial data \eqref{e:indat} differs from the
intended quasi-breather configuration.  In general, using it as initial
data, first a small portion of the energy is quickly emitted by
radiation, and then a very long living localized oscillating
configuration remains. However, generally, the frequency and amplitude
of this ``unclean'' oscillon state also possesses a lower frequency
modulation. We observed that the amplitude of this modulation can be
significantly decreased by multiplying the initial data by an
appropriate constant. By making a one-parameter tuning code we were
able to obtain oscillon states clean enough for studying their basic
energy loss rate. Otherwise one could not distinguish between the
energy emitted by the oscillon and the energy released by the decay of
the low frequency modulation. 

Through the whole domain of simulation the sum \eqref{e:indat} with
$N=3$ proved to yield the cleanest oscillon states.  It is worth
mentioning, that even for quite large values of $\varepsilon$, after
the tuning the asymptotic series yields appropriate initial data,
although the $\varepsilon$ value of the initial data and the one
calculated from the frequency by $\varepsilon=\sqrt{1-\omega^2}$
during the time evolution may differ.

\begin{figure}[!ht]
    \begin{center}
    \includegraphics[width=10cm]{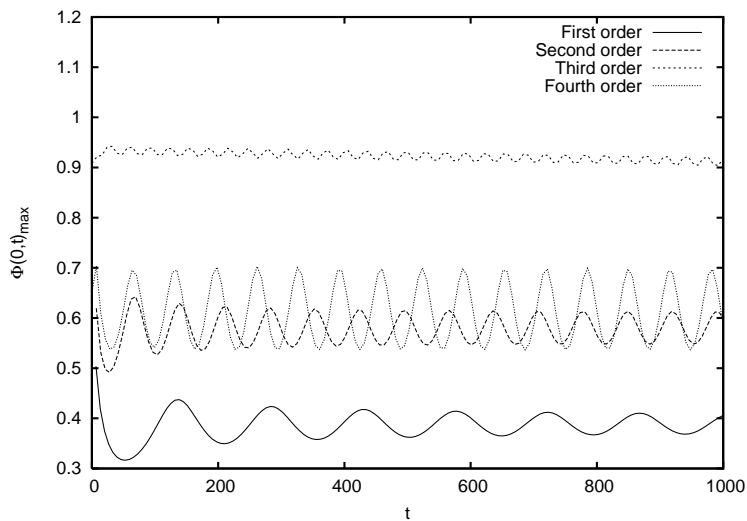}
    \end{center}
    \caption{Evolution from initial data in various orders for
      $\varepsilon=0.6$ in the $\phi^4$ theory; field value maximums at the center are
      plotted. \label{f:ord}}
\end{figure}

\begin{figure}[!ht]
    \begin{center}
      \subfigure[$\varepsilon$ calculated from the
      frequency]{\includegraphics[width=7cm]{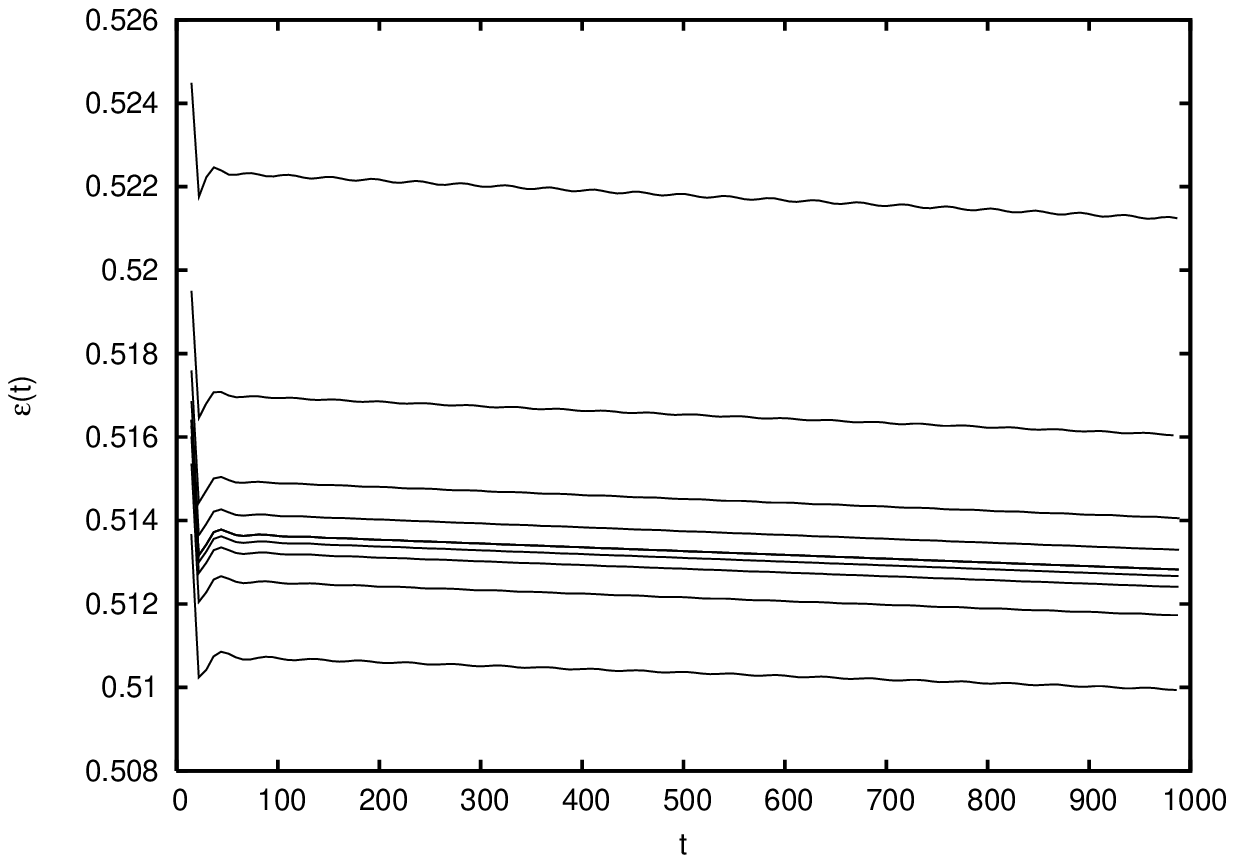}}
      \subfigure[Field value
      maximums]{\includegraphics[width=7cm]{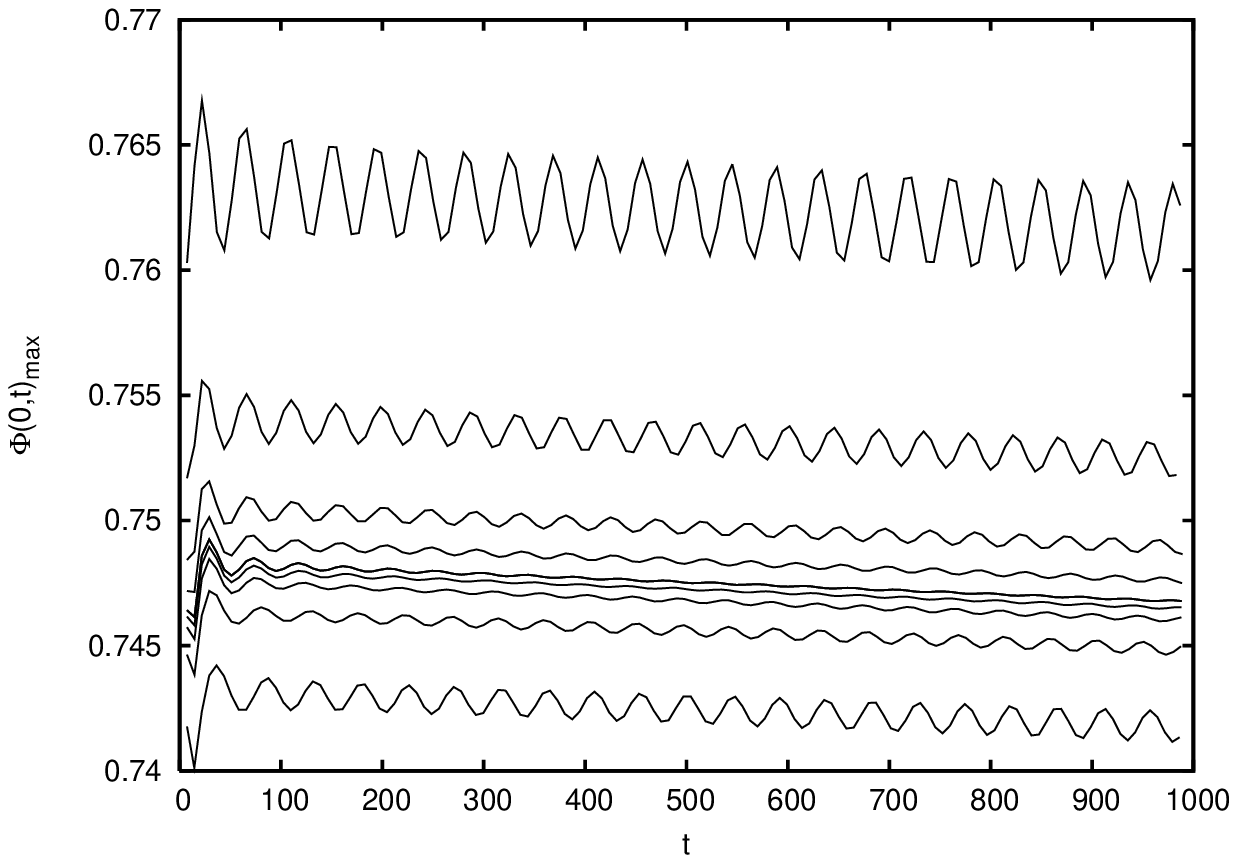}}
    \end{center}
    \caption{Time evolution of $\varepsilon$ and central amplitude for
      tuned $\varepsilon=0.5$ initial data in the $\phi^4$ theory. Multiplication factors are
      in the range $[0.9978,1.0169]$, and the smoothest evolution is with
      $1.00263$.} \label{f:tun}
\end{figure}

We illustrate the main steps by the example of the $\phi^4$
theory. The first few terms of the asymptotic series are:
\begin{eqnarray}
\phi_1^{(\tau=0)}&=&\sqrt{\frac{2}{3}}\, S\nonumber\\
\phi_2^{(\tau=0)}&=&\frac{1}{3}S^2\nonumber\\
\phi_3^{(\tau=0)}&=&\frac{1}{9}\sqrt{\frac{2}{3}}
\left(-\frac{25}{2}S^3+\frac{49}{2}S\right)\label{e:indatphi4}\\
\phi_4^{(\tau=0)}&=&\frac{1}{9}\left(-\frac{125}{6}S^4
+\frac{103}{3}S^2\right) \,.\nonumber
\end{eqnarray}
The most naive estimate for an ideal truncation of an asymptotic
series is that we should find the order where the terms of the series
are equally big and higher order terms are starting to grow from this
threshold. From \eqref{e:indatphi4} for $0.3<\varepsilon<0.6$
truncation of the series at third order appears to be
appropriate. This is also supported by the results of the numerical
simulations.  In the $\phi^4$ theory we found that $\varepsilon=0.6$
oscillons are the biggest ones which we can clean from the noise and
for which the adiabatic hypothesis works. Fig. \ref{f:ord} shows the
obtained ``unclean'' oscillon states from various order initial data. 
It is apparent that the order $N=3$ gives the state with the smallest
amplitude modulation. As already noted, multiplying the initial data
with a constant close to $1$ decreases the amplitude of this
modulation even more.  Figure \ref{f:tun} shows how effectively this
method smooths the oscillon for $\varepsilon=0.5$.

\subsection{Reliability of the numerical results concerning oscillon
  radiation\label{sec:reliability}}

The reliability of the subsequent results concerning oscillon
radiation can be estimated by checking how well our code simulates the
exactly periodic sine-Gordon breather.  We use the breather field
configuration \eqref{e:sG-breather} as initial data for the time
evolution. We note that using the series \eqref{e:sG-series} truncated
at third order and employing the tuning method gives the same order of
radiation. Table \ref{t:sg}. contains for different
resolutions the radiation power
and $W/E_B$ where $E_B$ is the total
energy of the breather ($E_B=16\varepsilon$). The number of lattice sites are given by the formula
$N_{lat}=RES\times 128$.
\begin{table}[!ht]
\vspace{0.5cm}
\begin{center}
\begin{tabular}{|c|c|c|}
\hline
$RES$  & $W$ & $\left|\frac{W}{E_B}\right|$\\
\hline
4 & $-1.18062\cdot10^{-6}$ & $1.48\cdot10^{-7}$\\
8 & $-3.66652\cdot10^{-8}$ & $4.58\cdot10^{-9}$\\
16 & $-1.14045\cdot10^{-9}$ & $1.42\cdot10^{-10}$\\
32 & $-3.5542\cdot10^{-11}$ & $4.44\cdot10^{-12}$\\
64 & $-1.10946\cdot10^{-12}$ & $1.38\cdot10^{-13}$\\
\hline								
\end{tabular}
\end{center}
\caption{Resolution dependence of radiation power $W$ in the sine-Gordon
  theory. The third column gives the ratio compared to the energy of
  the breather $E_B$. The number of lattice sites are $N_{lat}=RES\times
  128$. \label{t:sg}}
\end{table}
It can be observed that the duplication of the lattice points results
in the decrease of radiation due to lattice effects with a factor of
30. From now on we will keep ourselves to the following rule: we only
consider oscillons at a certain resolution if the numerically
calculated $W/E$ is at least one magnitude bigger than the value
given in Table \ref{t:sg}. If this condition fails, that means one
has to use higher number of lattice sites to measure the radiation rate
reliably, which, however, can make the simulation time impracticably
long. This is the reason why the energy loss of oscillons for very
small $\varepsilon$ cannot be obtained by our method.

\subsection{Oscillons in the $\phi^4$ theory\label{sec:phi4}}

\subsubsection{Confirmation of the adiabatic hypothesis and the radiation law}

In Table \ref{adat} we give the energy loss for the $\phi^4$ theory
for the biggest possible interval where we are able to use our
numerical approach. 
\begin{table}[!ht]
\vspace{0.5cm}
\begin{center}
\begin{tabular}{|c|c|c|c|} 
\hline
$\varepsilon$  & $W$ & $\left|\frac{W}{E}\right|$ & $\nu_2$\\ 
\hline
$0.5858$	&$-1.05\cdot 10^{-5}$	&$1.34\cdot 10^{-5}$	&$0.0905$ \\ 
$0.58255$	&$-9.58\cdot 10^{-6}$	&$1.23\cdot 10^{-5}$	&$0.0887$ \\ 
$0.5778$	&$-8.45\cdot 10^{-6}$	&$1.10\cdot 10^{-5}$	&$0.0866$ \\ 
$0.57123$	&$-7.12\cdot 10^{-6}$	&$9.35\cdot 10^{-6}$	&$0.0839$ \\ 
$0.56261$	&$-5.69\cdot 10^{-6}$	&$7.58\cdot 10^{-6}$	&$0.0807$ \\ 
$0.55183$	&$-4.27\cdot 10^{-6}$	&$5.80\cdot 10^{-6}$	&$0.0768$ \\ 
$0.53894$	&$-3.01\cdot 10^{-6}$	&$4.18\cdot 10^{-6}$	&$0.0725$ \\ 
$0.52418$	&$-1.98\cdot 10^{-6}$	&$2.84\cdot 10^{-6}$	&$0.0679$ \\ 
$0.49217$	&$-7.61\cdot 10^{-7}$	&$1.16\cdot 10^{-6}$	&$0.059$ \\ 
$0.48278$	&$-5.64\cdot 10^{-7}$	&$8.77\cdot 10^{-7}$	&$0.0565$ \\ 
$0.47334$	&$-4.14\cdot 10^{-7}$	&$6.57\cdot 10^{-7}$	&$0.0542$ \\ 
$0.46387$	&$-3.01\cdot 10^{-7}$	&$4.87\cdot 10^{-7}$	&$0.0520$ \\ 
$0.45437$	&$-2.17\cdot 10^{-7}$	&$3.58\cdot 10^{-7}$	&$0.0498$ \\ 
$0.44485$	&$-1.54\cdot 10^{-7}$	&$2.60\cdot 10^{-7}$	&$0.0478$ \\ 
$0.43533$	&$-1.09\cdot 10^{-7}$	&$1.87\cdot 10^{-7}$	&$0.0459$ \\ 
$0.42581$	&$-7.58\cdot 10^{-8}$	&$1.34\cdot 10^{-7}$	&$0.0441$ \\ 
$0.41629$	&$-5.23\cdot 10^{-8}$	&$9.43\cdot 10^{-8}$	&$0.0424$ \\ 
$0.40679$	&$-3.58\cdot 10^{-8}$	&$6.59\cdot 10^{-8}$	&$0.0408$ \\ 
$0.3973$	&$-2.42\cdot 10^{-8}$	&$4.57\cdot 10^{-8}$	&$0.0394$ \\ 
$0.35954$	&$-4.13\cdot 10^{-9}$	&$8.61\cdot 10^{-9}$	&$0.0334$ \\ 
$0.32017$	&$-4.45\cdot 10^{-10}$	&$1.04\cdot 10^{-9}$	&$0.0278$ \\ 
$0.28404$	&$-4.40\cdot 10^{-11}$	&$1.16\cdot 10^{-10}$	&$0.0258$ \\ 
\hline								
\end{tabular}
\end{center}
\caption{Radiation power $W$ and $\varepsilon$ in the $\phi^4$ theory 
  from various initial data. For the oscillons $\varepsilon$ is measured from the frequency during time evolution. The $\left|W/E\right|$ column determines the resolution $RES$ required for the simulation, while $\nu_2$ is calculated from Eq. \eqref{e:radlow1}.\label{adat}}
\end{table}
The lower bound is limited by the need of time to
perform high resolution simulations; we can see only lattice effects
below the lowest $\varepsilon$ values in the table. Above the upper
bound the oscillon decay is very fast and we cannot determine $W$ and
$\varepsilon$ reliably. We will see the radiation law fail for large
$\varepsilon$, in these cases the interaction with the radiation field
may become essential, and the system does not evolve through
undistorted quasi-breather states. The $\nu_2$ values in Table \ref{adat} are calculated from the measured radiation power and $\varepsilon$ by using the theoretical radiation law \eqref{e:radlow1}. 

\begin{figure}[!ht]
    \begin{center}
    \subfigure[The theoretical and semi-empirical radiation law .]{\includegraphics[width=7cm]{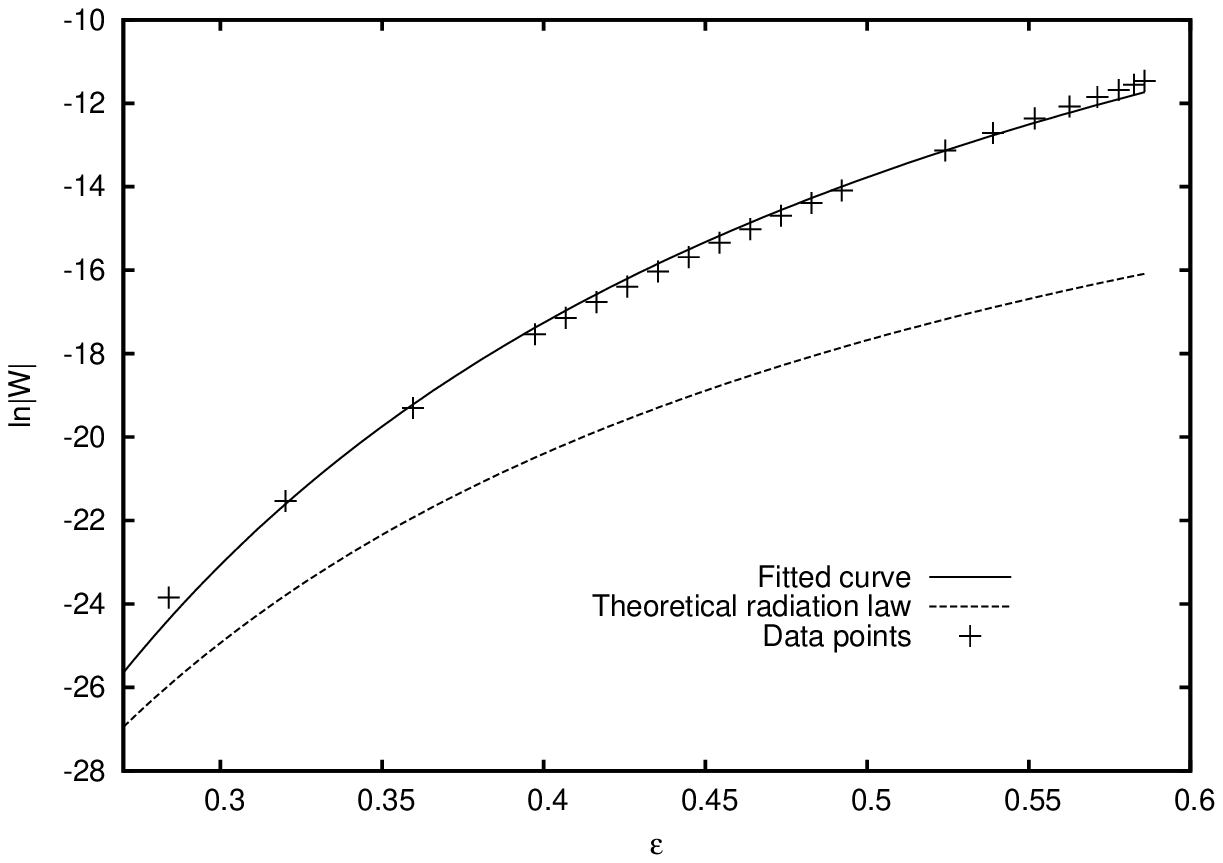}}
    \subfigure[$\nu_2$ as a function of $\varepsilon$.]{\includegraphics[width=7cm]{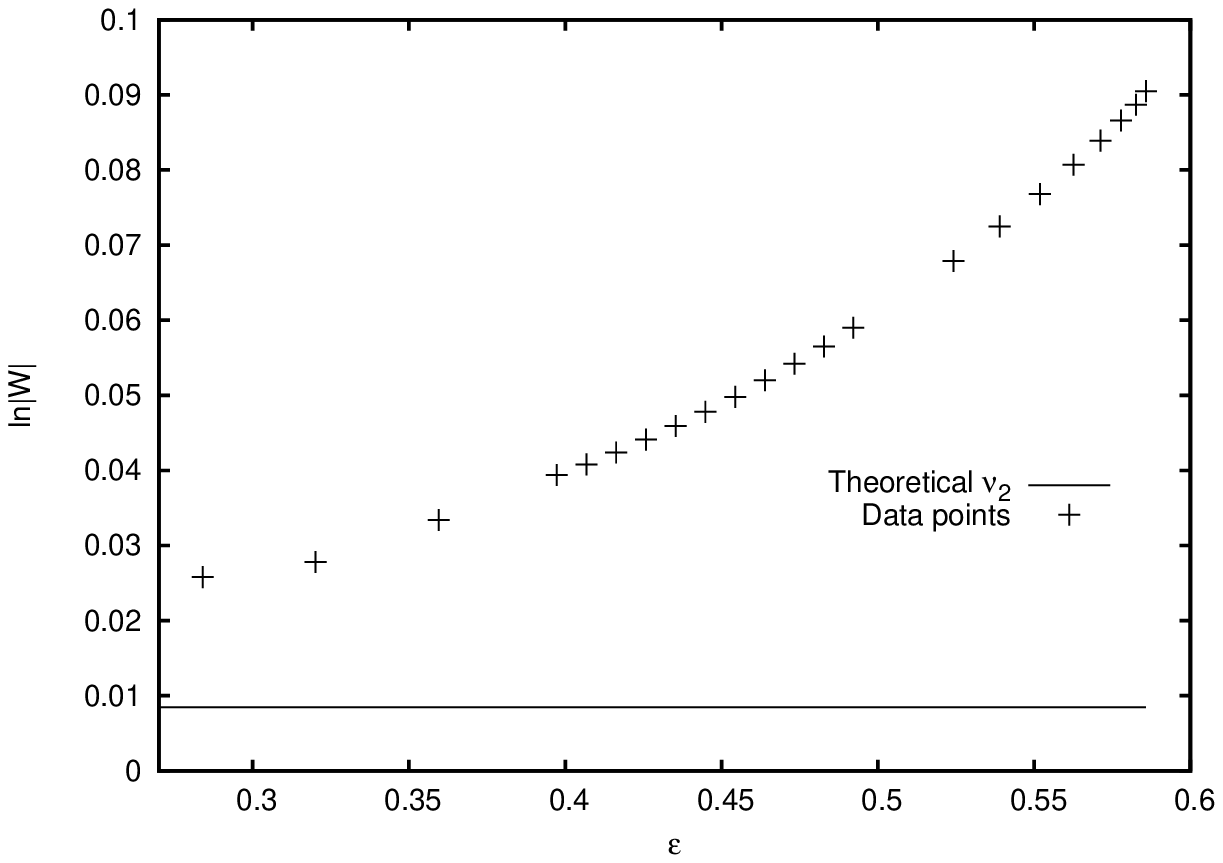}}
    \end{center}
    \caption{The measured data points, the theoretical radiation law and the results of the first fit with parameters to be found in Eq. \eqref{e:fitall} in the case of the $\phi^4$ theory. We see the points are approaching the theoretical curve as $\varepsilon$ decreases.
      \label{fit}}
\end{figure}

We intend to confirm the radiation law, Eq. \eqref{e:radlow1} numerically.
We performed two fits with two free parameters on the logarithm of the
data shown on Figure \ref{fit}(a). In the first fit we took into account all the data points, while in the second we fitted for data points with $\varepsilon<0.42$ to get closer to the theoretical pole term. We define the semi-empirical radiation law: 
\begin{equation}
\frac{\mathrm{d} E}{\mathrm{d}t}
=W=-8\sqrt{3}\;\nu_2^2\cdot\exp\left[-\frac{\sqrt{3}\;\pi\,b}{\varepsilon}\right]\,,\label{e:fit1}
\end{equation}
where $\nu_2$ and $b$ are parameters to be fitted. Eq. \eqref{e:fit1} is to be interpreted as follows. For finite values of $\varepsilon$ there are various higher order corrections to the radiation law which are now encoded in two effective parameters, $\nu_2$ and $b$. These corrections originate in Eq. \eqref{e:om2}. We remark that the value of $b_{theory}$ will increase accordingly. When fitting for all the data points the difference between $b$ and $b_{theory}=1$ is bigger, than in the case of the fit for $\varepsilon<0.42$ points, because for smaller $\varepsilon$ values the correction due to other pole terms are smaller.   

If we set $b=b_{theory}$ in Eq. \eqref{e:fit1} we get back the theoretical radiation law, Eq. \eqref{e:radlow1}. On Figure \ref{fit}(b) we plotted the $\nu_2$ values from Table \ref{adat}. We see that we are getting closer to the theoretical $\nu_2$ as $\varepsilon$ decreases. In Eqs. \eqref{e:nu1}, \eqref{e:nu2} we give the results of the fits of $\nu_2$ with $b=b_{theory}$ on all the data points and for data points with $\varepsilon<0.42$ respectively.

The result of the first fit supports the functional dependence got from the theoretical arguments, while the second fit leads us closer to the theoretical values of the parameters:
\begin{align} 
	\nu_{2,fit}&=0.29\pm0.03 & b_{fit}&=1.28\pm0.2 &&\text{for all data points,} \label{e:fitall}\\
	\nu_{2,fit}&=0.12\pm0.02 & b_{fit}&=1.17\pm0.2 && \text{for data points with $\varepsilon<0.42$,}\\
	\nu_{2,fit}&=0.054\pm0.004 & b&=1 &&\text{for all data points,} \label{e:nu1}\\
	\nu_{2,fit}&=0.034\pm0.004 & b&=1 && \text{for data points with $\varepsilon<0.42$,} \label{e:nu2}\\
	\nu_{2,theory}&=0.00845 & b_{theory}&=1\;.
\end{align}

Unfortunately the theoretical $\nu_2$ value of the $\phi^4$ theory is exceptionally small. This might be the reason why our data are consistent with the theoretical pole terms and why there is a difference in the value of $\nu_2$. It should be noted that we are fitting over 7 orders of magnitude. An other independent source of discrepancy could be the tuning method and that the superfluous energy from the oscillon core goes out in shells \cite{fodorkg}. These shells are observed to be present even for large time and their movement could increase the radiation power measured by our method.

Let us turn our attention to the verification of the semi-empirical radiation law determined from the short-time evolution of various oscillon states. We shall confirm from a long-time simulation, that the evolution from an initial oscillon state is driven by our radiation law. We claim two states to be identical if they posses the same $\varepsilon$ value with the same energy for a relatively long time. This should mean that the field configurations are very close to each other, the interaction with the radiation field is negligible. By testing these properties in two different simulations we can match these data and get a longer process.

To compare the time dependence of the measured energy with the
predicted one, we need the function $E(\varepsilon)$. This should not be
a problem, as we can measure this function from the short simulations. The results of simulations are in very good agreement with the semi-empirical radiation law. We would like to emphasise that the radiation law is not a simple fit, it has been determined independently; we only use the initial $\varepsilon$ value and the function $E(\varepsilon)$ from the simulation to get the predicted curve.

\begin{figure}[!ht]
    \begin{center}
    \subfigure[The energy]{\includegraphics[width=10cm]{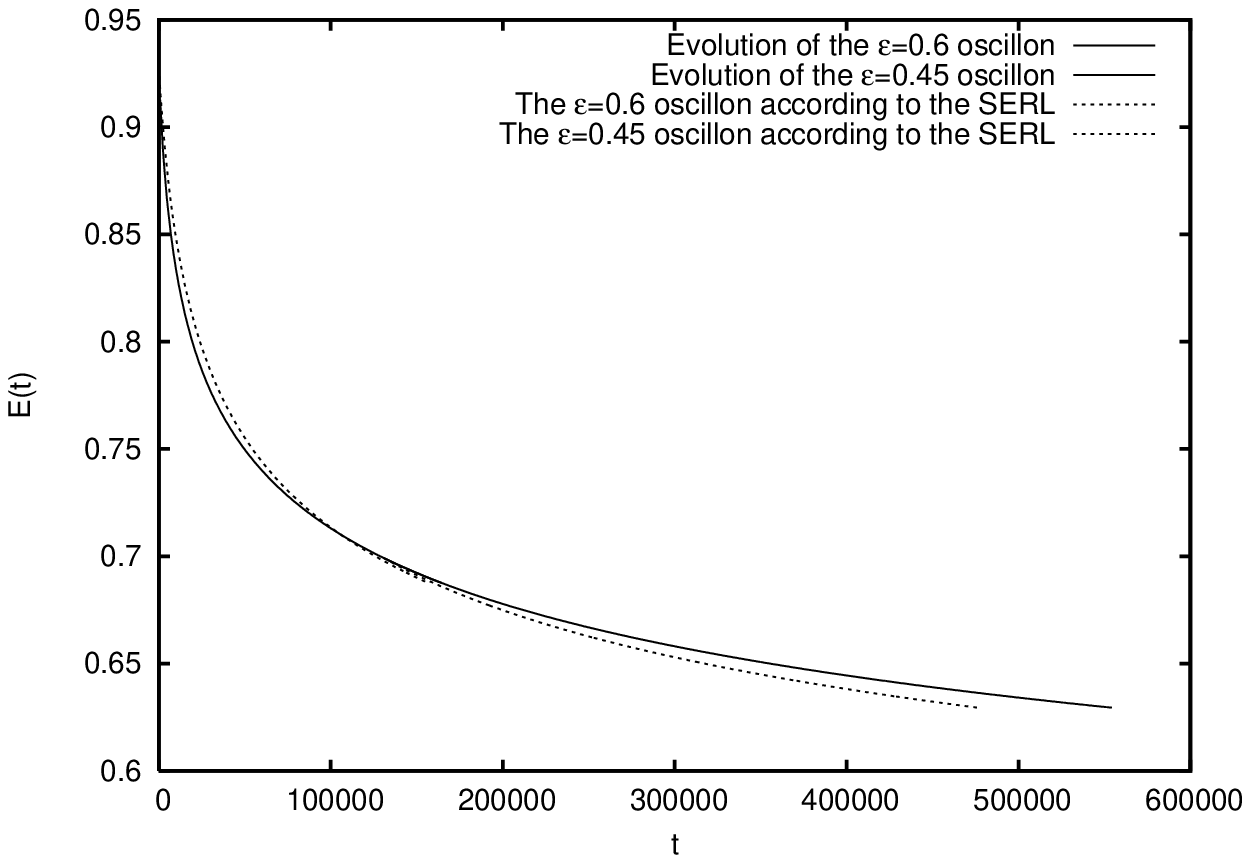}}
    \subfigure[$\varepsilon$ versus time]{\includegraphics[width=10cm]{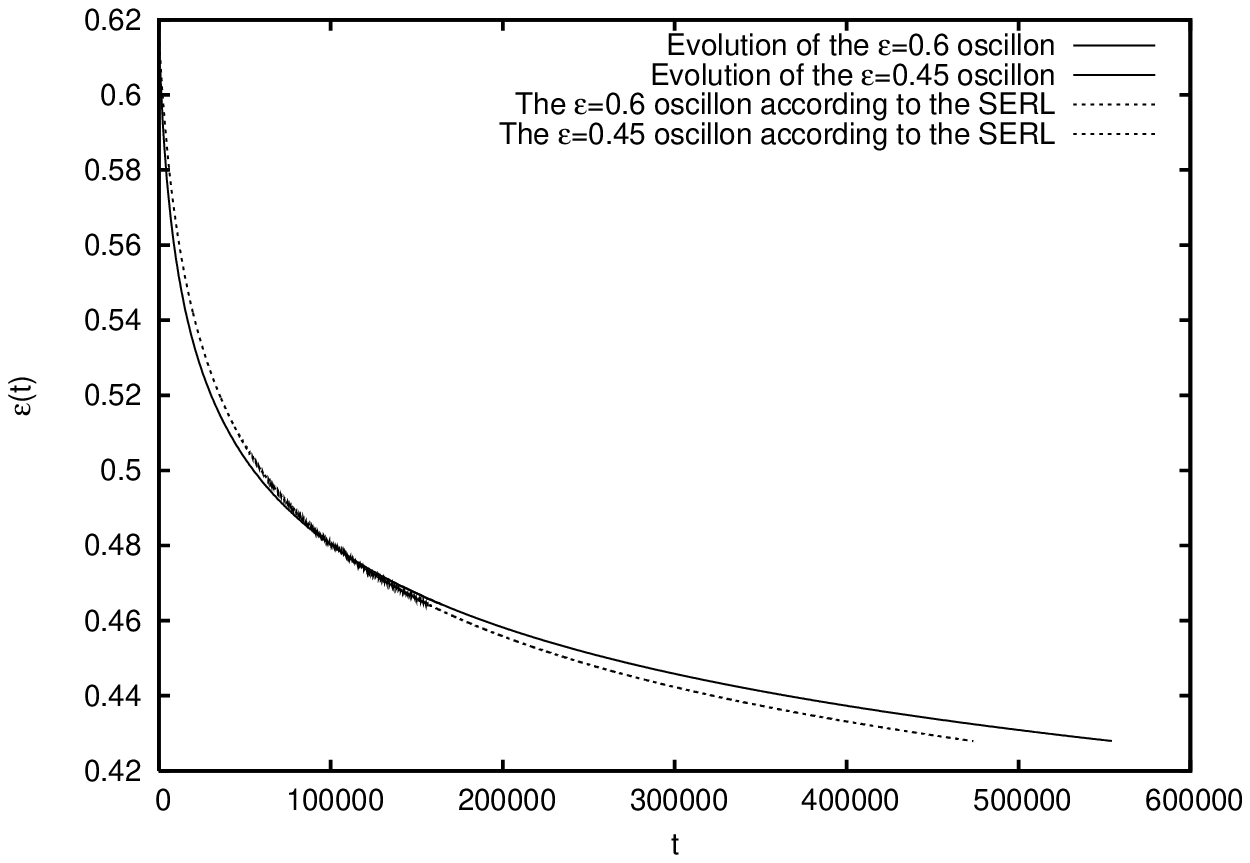}}
    \end{center}
    \caption{The evolution of an $\varepsilon=0.6$ and an $\varepsilon=0.45$ oscillon in the $\phi^4$ theory matched together; both the numerical data and the prediction of the SERL are plotted. The time evolution of the two oscillons are overlapping in a long time interval around $t=100000$.} \label{jo}
\end{figure}

On Figure \ref{jo} we see an $\varepsilon=0.6$ oscillon evolving. We
used two simultaneously performed simulation for this graph (starting
from $\varepsilon=0.6$ and $\varepsilon=0.45$), which perfectly fit
together; another fact backing the hypothesis of adiabatic
evolution. The solid lines follow the evolution observed in numerical
simulations, while the dashed lines are the predictions of the
semi-empirical radiation law (SERL). We end both curves at the same
$\varepsilon$ value. 


\subsubsection{Oscillons from kink-antikink initial data}

After being able to create and examine clean oscillon states we aim to identify, what the logarithmically decaying object is in our terminology that Geicke found in Refs.\ \cite{geicke1,geicke2}. It turns out that the objects he found are composite oscillon states and there is a continuum of them, without any essential difference between these configurations in contradiction with what he conjectured. These objects do not obey the semi-empirical radiation law set for clean oscillon states in their early but very long period ($\approx 150000$) of life, they radiate more rapidly, as for the energy stored in the modulation modes of frequency has different radiation properties. However, their lifetime is in the same magnitude of oscillons with approximately the same frequency and energy. After their initial stage of evolution they begin to obey the semi-empirical radiation law, their modulation, however, does not disappear. We shall confirm the logarithmic fits of Ref.\ \cite{geicke1} and in this stage of observation we can explain qualitatively what he found and how his results are related to the analytic considerations.

\begin{figure}[!ht]
    \begin{center}
    \includegraphics[width=10cm]{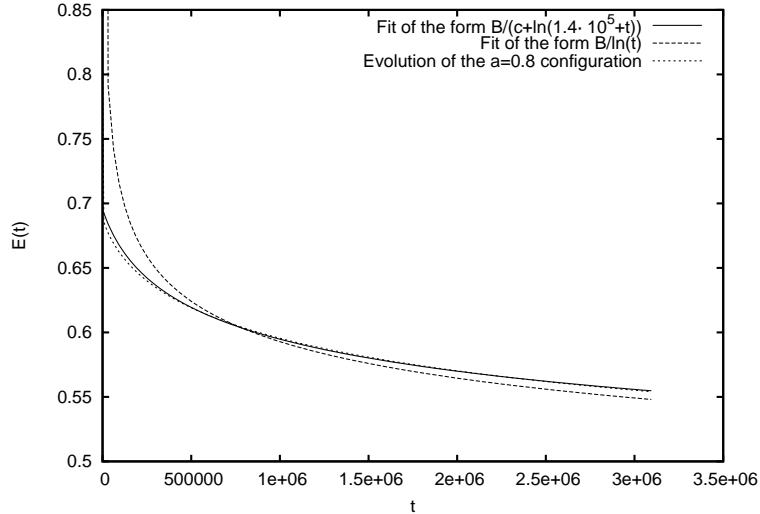}
    \end{center}
    \caption{The evolution of the kink-antikink initial data in the $\phi^4$ theory with $a=0.8$ parameter illustrated by the energy in the entire simulation.\label{geicke2}}
\end{figure}

Following Ref.\ \cite{geicke1}, we examine a kink-antikink pair initially at rest:
\begin{align}
     \phi(x,t=0)&=\tanh\left[\frac{x-a}{2}\right]-\tanh\left[\frac{x+a}{2}\right]+1\\
    \partial_t \phi(x,t=0)&=0\;.
\end{align}
For different values of $a$ we examined the evolution of the initial state. The resolution was selected low ($RES=4$) because we were not interested in the exact decay rate, we aimed to draw qualitative consequences. The results are collected in Fig. \ref{geicke2}.
\begin{figure}[!ht]
    \begin{center}
    \includegraphics[width=10cm]{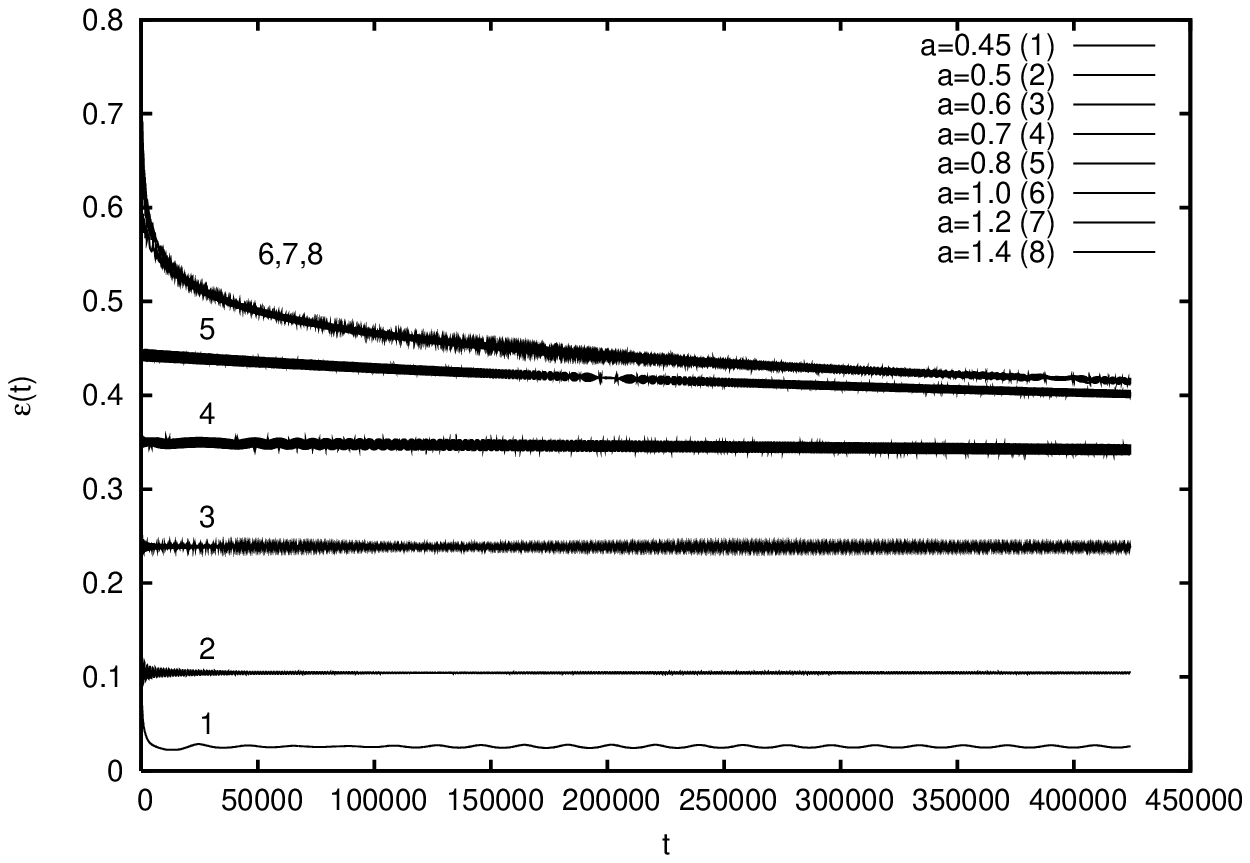}
    \end{center}
    \caption{The evolution of the kink-antikink initial data in the $\phi^4$ theory for various $a$ values; $\varepsilon$ versus time plotted.}
\end{figure}
For bigger values of $a$ ($a>1.0$) we find, that after an early period of formation approximately the same state arises. For moderate values of $a$ we see different states arising in a relatively short period of time; these states radiate very slowly. Below a certain value of $a$ ($a\lessapprox 0.45$) we do not observe any definite oscillon core. This behaviour is remarkable: we found the same pattern in the collision of a sine-Gordon soliton-antisoliton pair initially at rest, the same state arises for $a>1.4$ and for the region $a\lessapprox 0.8$ we see no definite oscillon state.

To find out how these configurations radiate we ran a simulation with $RES=8,\;a=0.8$; this is the initial data of Geicke in Ref.\ \cite{geicke1}. The following figure shows us what the semi-empirical radiation law would predict and what really happens. The lumps emerging from kink-antikink initial data radiate faster then a clean oscillon due to its frequency modulation. The following asymptotic logarithmic fit considered in Ref.\ \cite{geicke1} works well:
\begin{align}
    E(t)&=\frac{B}{c+\ln(t+\sqrt2\cdot10^5)}\\
    c&=0.52\pm0.12 & B&=8.60\pm0.07\;.
\end{align}
Ref.\ \cite{geicke1} uses $m=\sqrt2$, its result for the $B$ parameter in the units we use in this paper is $B_g=9.05$, the reason for the difference between $B$ and $B_g$ might be the use of Sommerfeld boundary condition in its simulation instead of compactification. We fitted the function which we get by extrapolating the theoretical result for infinitesimal, Eq. \eqref{e:energy-time} oscillons to our case as well:

\begin{align}
    E(t)&\approx\frac{B}{\ln t}\\
    B_{theory}&=\frac{4\,\pi}{\sqrt 3}\approx 7.255\\
    B_{fit}&=8.190\pm0.007\;.
\end{align}
We see that the numerical values from the fit and from theory are satisfactory close to each other. We note that such a logaritmical fit is not influenced by the value of $\nu_2$ in the theoretical radiation law, Eq. \eqref{e:radlow1}. (Ref.\ \cite{geicke1} claims that the results of its fit, $B=14.395$ agrees with the theoretical result $B=14.503$ in its units; however, the correct $B$ value from theory with $m=\sqrt2$ mass scale is $B=10.260$.)

We conclude that the initial state observed by Ref.\ \cite{geicke1} evolves as a
complex oscillon-like object. After the 'early' period of $t\approx
150000$ it looses a large part of its energy and an oscillon with modulated frequency is created. Through its evolution the
$\varepsilon$ value changes from 0.45 to 0.38. From the solutions of the
semi-empirical radiation (SERL) law it can be clearly seen that the
initial configuration decays much faster than an oscillon, however if
we regard the $t\approx 150000$ as initial data, it obeys the radiation
law with high precision despite the presence of modulation. (We did
not plot the prediction of the SERL with $t_0=150000$ in Figure
\ref{geicke2}, as the difference between the prediction and the
numerical simulation results do not differ visibly.) It seems to us
that the modulation degree of freedom is an adiabatically decaying
mode.

\subsection{Verification of the theoretical radiation law\label{sec:verification}}

We will examine oscillons in the specific symmetric $\phi^6$ theory in which oscillon radiation is the largest, i.e.\ when $K$ is maximal. We determined the $g_5$ value for this theory in subsection \ref{ss:bs}. Thus, contrary to the $\phi^4$ theory the value of $K$ is maximal and the next to leading order corrections are significantly smaller compared to the leading order term, than in the case of the $\phi^4$ theory because the $\phi^6$ potential is symmetric. These corrections originate in \eqref{e:om3}. It should be kept in mind that because of the smaller pole term we cannot go as low in $\varepsilon$ as in the $\phi^4$ case. We collect the results of the simulations in Table \ref{prec} in analogy to Table \ref{adat}:
\begin{center}
\begin{table}[!htbp]
\vspace{0.5cm}
\begin{center}
\begin{tabular}{|c|c|c|c|}
\hline
$\varepsilon$  & $W$ & $\left|\frac{W}{E}\right|$ & $\nu_3$\\
\hline
$0.391886$  &$-1.256\cdot 10^{-8}$  &$1.20\cdot 10^{-8}$   &$1.614$ \\
$0.389854$  &$-7.983\cdot 10^{-9}$  &$7.68\cdot 10^{-9}$   &$1.365$ \\
$0.366842$  &$-5.692\cdot 10^{-10}$ &$5.82\cdot 10^{-10}$  &$0.745$ \\
$0.352178$  &$-5.483\cdot 10^{-10}$ &$5.84\cdot 10^{-10}$  &$1.211$ \\
$0.339333$  &$-3.291\cdot 10^{-10}$ &$3.64\cdot 10^{-10}$  &$1.512$ \\
$0.333239$  &$-2.225\cdot 10^{-10}$ &$2.50\cdot 10^{-10}$  &$1.579$ \\
$0.304623$  &$-2.293\cdot 10^{-11}$ &$2.82\cdot 10^{-11}$  &$1.774$ \\
\hline
\end{tabular}
\end{center}
\caption{Radiation power for oscillons with different $\varepsilon$ values in the symmetric $\phi^6$ theory. } \label{prec}
\end{table}
\end{center}

We performed two fits: in the first we fitted both $\nu_3$ and $b$, in the second we used $b=b_{theory}$ and fitted $\nu_3$ as we were interested in how accurately we could determine the value of $\nu_3$ from numerical simulations.
\begin{align}
    \frac{\mathrm{d} E}{\mathrm{d} t}&=-24\sqrt2\,\nu_3^2\,\exp\left[-\frac{\sqrt8 \pi\,b}{\varepsilon}\right]\\
    \nu_{3,fit}^{(1)}&=0.4\pm1.2 & b_{fit}^{(1)}&=0.89\pm0.1\\
    \nu_{3,fit}^{(2)}&=1.35\pm0.1 \\
    \nu_{3,theory}&=0.9098 & b_{theory}&=1\;.
\end{align}
\begin{figure}[!htbp]
    \begin{center}
    \includegraphics[width=12cm]{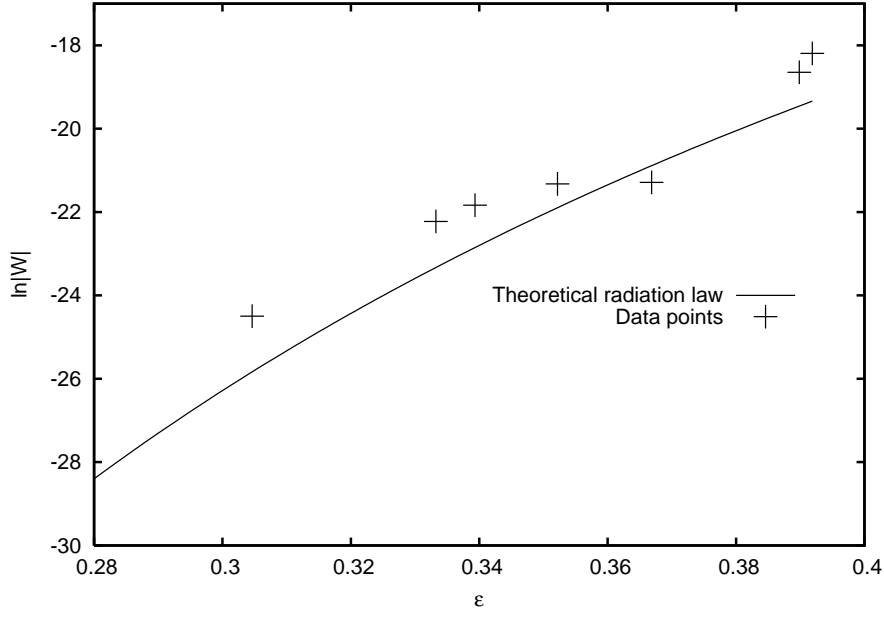} 
    \caption{The relation between the numerical data and the theoretical radiation law in the $\phi^6$ theory. We plotted the data points and the theoretically predicted curve only, as the results of the fits would only make the graph less understandable.\label{f:phi6}}
    \end{center}
\end{figure}

One can notice an anomalous point among the data points in Fig. \ref{f:phi6}.
This anomalous configuration was further examined by us, but no visible anomaly
in the field configuration was observed. The oscillon has a proper oscillating
tail and a smooth core. Its radiation power was measured for different
resolutions and also by using different values for the conformal factor
$\kappa$, but all these had no substantial effects. As it is, we do not know
whether this anomaly is an awkward lattice effect or genuine one i.e.\ present
in the continuum limit as well.

\begin{figure}[!htbp]
    \begin{center}
    \includegraphics[width=12cm]{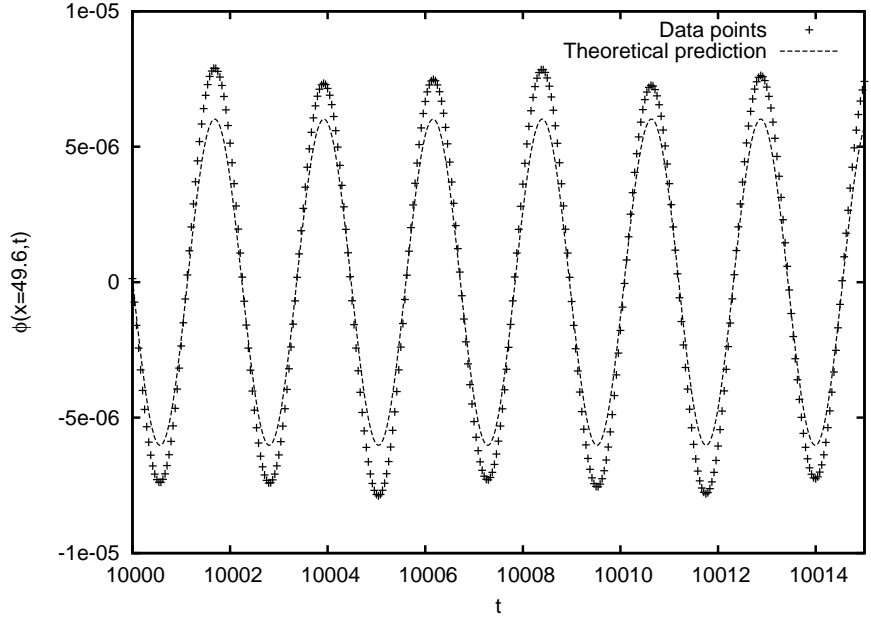} 
    \caption{The field configuration observed in numerical simulations and the theoretically predicted wave amplitude for $\varepsilon=0.352178$ oscillon in the $\phi^6$ theory.\label{f:radfield}}
    \end{center}
\end{figure}

We would like to investigate the problem of oscillon radiation from
the perspective of field configuration as well. In
Fig. \ref{f:radfield} we show the field values as a function of time
at $x=49.6$ for the $\varepsilon=0.352178$ oscillon. We plot the
theoretical prediction of this oscillating tail, so that the frequency
of the wave is set to $3\omega$, the amplitude is determined by theory
and only the phase of the wave is fitted. We find satisfactory
agreement in the case of wave amplitude and precise agreement in the
case of frequency. If we compare the radiation power calculated from
this plane wave and the one in Table \ref{prec} we find good
agreement. Hence we conclude that the oscillons on the lattice loose
energy via radiation and that out assumptions were correct when
deriving the radiation law for small amplitude oscillons.

There are various possible explanations for the discrepancy between theoretical and numerical results. Firstly,
we could argue that the $\varepsilon$ values used in numerical
simulations are too big and the theoretical calculation of the
radiation amplitude only works for infinitesimal $\varepsilon$
values. For finite $\varepsilon$ values we can only expect an
approximate agreement. Secondly, we cannot be sure about the initial
data. Although the frequency of our objects is very stable there is no
way to decide whether we work with undistorted oscillons. The
configuration with anomalous radiation emphasizes these
problems. Lattice effects are less probable to play a role, as we see
no major resolution and $\kappa$ dependence of the radiation powers.

We conclude that we can accurately determine the $b$ parameter of the radiation law from the numerical simulation of oscillon decay in the case of both the $\phi^4$ and $\phi^6$ case. The parameter $\nu_2$ can be less accurately determined in the case of the $\phi^4$ theory. There the next to leading corrections play a role because of the finite $\varepsilon$ oscillons and the value of $\nu_2$ is too small to be seen. In the case of the $\phi^6$ theory the next to leading order corrections are smaller and the value of $\nu_3$ is bigger, hence we can get the value of $\nu_3$ from the simulations with satisfactory precision. The investigation of the radiation field of the oscillons show equally good agreement with the theoretical formulae giving a firm basis for the theoretical calculations from another perspective.

\section{Conclusions}\label{s:conc}
In a general class of one dimensional scalar field theories
we have computed the magnitude of the radiative tail of oscillons,
determining their energy loss, in the small amplitude limit. 
The magnitude of the tail is non-perturbatively small in the amplitude.
We have used the Segur-Kruskal method of matched asymptotic expansion
together with Borel summation techniques to calculate it.
These results have also been verified numerically. 
We have also performed numerical simulations to compute directly
the energy loss of oscillons, as well as the radiative tail.
The numerical results are in a satisfactory agreement with the theoretical predictions. 

\section{Acknowledgments}

This research has been supported by OTKA Grants No. K61636,
NI68228.

\end{document}